\documentstyle[prl,aps,preprint,tighten,floats,aps,epsf,psfig]{revtex}

\def\be{\begin{equation}}
\def\ee{\end{equation}}
\def\bear{\begin{eqnarray}}
\def\eear{\end{eqnarray}}
\def\beqn{\begin{eqnarray}}
\def\eeqn{\end{eqnarray}}

\def\re{r}

\def\sp{\sigma^{+}}
\def\sm{\sigma^{-}}
\def\NS{Neveu-Schwarz }
\def\fer{\psi}
\def\bfer{{\bar\psi}}

\def\beq{\begin{equation} }
\def\eeq{\end{equation} }
\def\beqn{\begin{eqnarray} }
\def\eeqn{\end{eqnarray} }
\def\nolabel{\nonumber }
\def\ov{\overline}

\def\ran{\rangle}
\def\lan{\langle}

\def\lra{\leftrightarrow}
\def\rar{\rightarrow}

\def\e{{\rm e}}
\def\exp{{\rm e}}
\def\dc{{d^c}}
\def\uc{{u^c}}
\def\ec{{e^c}}
\def\vc{{\nu^c}}
\def\w{\omega}

\def\phm{\phantom{-}}

\def\beq{\begin{equation} }
\def\eeq{\end{equation} }
\def\beqn{\begin{eqnarray} }
\def\eeqn{\end{eqnarray} }

\def\be{\begin{equation} }
\def\ee{\end{equation} }
\def\ben{\begin{eqnarray} }
\def\een{\end{eqnarray} }

\def\nolabel{\nonumber }

\def\mod#1{{\rm (mod~2)} }

\def\Tr{{\rm Tr}}
\def\tr{{\rm tr}}

\def\Mp{M_{\rm P} }

\def\MS{M_{\rm string} }

\def\mh{-{1\over 2}}
\def\ph{ {1\over 2}}

\def\isqrt{{1\over \sqrt{2}}}

\def\half{{1\over 2}}

\begin{document}
\draft
\preprint{\vbox{\baselineskip=12pt
\rightline{CERN-TH/98-154}  
\vskip0.2truecm
\rightline{UPR-0784-T}
\vskip0.2truecm
\rightline{IEM-FT-173/98}  
\vskip0.2truecm
\rightline{hep-th/9805133}}}
\date{\today}
\title{Flat Directions in Three-Generation Free-Fermionic String Models} 
\author{Gerald Cleaver${}^{\dagger}$, Mirjam Cveti\v c${}^{\dagger}$,
Jose R. Espinosa${}^*$${}^{\dagger}$, Lisa Everett${}^{\dagger}$, and
Paul Langacker${}^{\dagger}$}
\address{${}^{\dagger}$Department of Physics and Astronomy \\
          University of Pennsylvania, Philadelphia PA 19104-6396, USA \\
${}^*$CERN, TH Division\\  
CH-1211 Geneva 23, Switzerland}
\maketitle
\begin{abstract}
In quasi-realistic string models that contain an anomalous $U(1)$
the non-zero Fayet-Iliopoulos term triggers the shifting of the original
vacuum to a new one along some flat direction, so that SUSY is preserved
but the gauge group is partially broken. The phenomenological study of
these models thus requires as a first step the mapping of the space of
flat directions. We investigate $F$- and $D$-flat
directions in  several three-generation $SU(3)_C\times SU(2)_L\times
U(1)_Y$  free-fermionic string models and discuss the typical scenarios
that generically arise. When they exist, we systematically construct
the flat directions that preserve hypercharge, only break Abelian
group
factors, and can be proven to remain $F$-flat to all orders in the
non-renormalizable superpotential.
\end{abstract}
\vskip2cm
\leftline{}
\leftline{CERN-TH/98-154}
\leftline{May 1998}
\pacs{}

\section{Introduction}
While the investigation of the phenomenology of string models continues to
face serious difficulties, such as the problem of the degeneracy of string
vacua and the absence of a satisfactory scenario for supersymmetry
breaking, classes of quasi-realistic models have been constructed that
warrant further phenomenological study.

The quasi-realistic string models are those which possess at least the
basic structure of the MSSM at the string scale ($M_{String}$).  Such
models have $N=1$
 supersymmetry,   the standard  model (SM) gauge group
  $SU(2)_L\times U(1)_Y\times SU(3)_C$ 
as a part of the
full gauge structure, and 
candidate fields for the three ordinary families and  at least two SM
Higgs doublets.  Classes of such models which satisfy the above
requirements have been constructed based on the weakly coupled heterotic
superstring. In particular, we focus on models based on the free-fermionic
construction \cite{ABKW,KLST,SW}.
In these constructions, the mass spectrum and
superpotential (in principle to all orders in the nonrenormalizable
terms) are calculable.

In addition to the standard model (observable) gauge group, the gauge
structure of
these models includes a non-Abelian (NAB)
hidden sector and a number of additional $U(1)$'s, of which 
at least one is generally 
anomalous\footnote{Conditions sufficient to keep all
$U(1)$'s anomaly free \cite{SW} have recently been discussed in \cite{KN} and
\cite{CF}, with an anomaly-free semi-GUT presented in \cite{CF}.} 
(by which we mean the charge trace over all matter states
is non-zero).
The appearance of such an anomaly 
will play a crucial role in the phenomenology of the model. 
The SM hypercharge is a linear combination of the
nonanomalous $U(1)$'s (or perhaps of the $U(1)$'s that can arise after the
non-Abelian hidden sector group is broken), which 
is chosen taking into account some basic phenomenological requirements.

In general, the particle spectrum is such that there are many
additional matter multiplets along with the MSSM particle content. 
Most of the states can
be classified according to those which are representations
of the observable sector NAB gauge group,  representations of the
hidden sector NAB gauge group, and NAB singlets (i.e., fields which are
singlets under NAB gauge groups but can carry $U(1)$ charges). However,
the
division between the observable and hidden sector gauge groups is tenuous
at best, because most of the fields are charged under the Abelian gauge
groups.  In some models, there also are ``mixed" states which are
 non-singlets under both the observable and hidden sector NAB gauge
groups.  Such states, if present, may have important consequences for the
phenomenology of the model.

The Green-Schwarz anomaly cancellation mechanism at genus-one in string
theory generates a constant Fayet-Iliopoulos (FI)
contribution  to the $D$- term of the anomalous $U(1)$ 
\cite{DSW,ADS,DKI,AS} which is proportional to the trace of the anomalous
charge over all of the fields in the model. 
The FI term would break supersymmetry in the original string vacuum,
but it triggers string-scale vacuum expectation values for certain scalar
fields. In the new shifted vacuum the $D$- and $F$-
flatness constraints are satisfied, supersymmetry is restored, and the
anomalous $U(1)$ is broken. There are many possibilities for this vacuum
shifting, and thus an analysis of the space of $D$- and
$F$- flat directions is the necessary first step before
addressing the phenomenology of the model.

In a previous paper \cite{CCEEL2}, 
we developed a method to classify in a systematic manner a subset of the
flat directions of a general perturbative string model with an anomalous
$U(1)$.  While there is no reason a priori why a flat direction should
not involve fields in non-trivial representations of NAB groups, we chose
to analyze 
those flat directions formed only from the NAB singlets  
for simplicity. Our
method involves determining a set of holomorphic gauge invariant
monomials (HIM's) \cite{LTGSM} that characterizes the moduli space of flat
directions under the 
nonanomalous $U(1)$'s. A straightforward
classification of the singlet fields of the model
according to their charges under a conveniently defined combination of
$U(1)$ charges (or of the HIM's according to their anomalous charges) 
can then determine by inspection if it is possible to
form a flat direction that can cancel the
anomalous $D$- term and preserve $U(1)_Y$. When such flat directions can
be formed, we presented a
systematic way to classify the subset of these flat directions which can
be   
proved to be 
$F$- flat to all orders in the nonrenormalizable superpotential
(and to all orders in string genus perturbation expansion).  
We demonstrated our method in \cite{CCEEL2} by applying it to a prototype
free-fermionic 
string model, Model 5 of Chaudhuri, Hockney and Lykken in \cite{CHL}, 
for which the results were particularly straightforward.

In this paper, we apply this method to a number of free-fermionic
three-generation string models, those presented in 
\cite{FNY1}, \cite{AF1}, and \cite{CHL}.  
We determine for each model whether it is possible to construct
hypercharge-preserving flat directions involving 
the NAB singlet fields which can cancel the FI $D$-term.  For the
models in which this is the case, we examine the space of such
flat directions in detail.

In Section II, we describe the construction and general properties of the
models under consideration.  We discuss our analysis of determining
whether one can 
construct good  flat directions 
and explain  
our procedure for determining a viable hypercharge in Section III.
In Section IV, we present the results for each model. The summary and conclusions
are given in Section V.

\section{$SU(3)_C\times SU(2)_L\times U(1)_Y$ Models of Perturbative 
Heterotic Strings}

\subsection{Standard Model Free-Fermionic Embeddings}

Quasi-realistic four-dimensional perturbative heterotic string models generally
contain gauge structures that extend beyond the rank four SM
$SU(3)_C\times SU(2)_L\times U(1)_Y$. 
In string models, gauge groups with ranks significantly 
larger than four are a generic by-product of conformal anomaly cancellation.   
The gauge structure of a stringy SM or (semi)-GUT can be expressed in the 
form,  
\beq
\{SU(3)_C\times SU(2)_L \in {\cal G} \}_{\rm obs}\times
{\cal G}^{\rm NA}_{\rm hid}\times 
\prod_n U(1)_n \times \prod_p {\Delta}_p\, , 
\label{ohs}
\eeq 
where ${\cal G}$ denotes a possible SM GUT or semi-GUT embedding,
${\cal G}^{\rm NA}_{\rm hid}$ contains the  hidden
sector NAB gauge factors, 
and ${\Delta}_p$ represents possible local discrete symmetries.
The various Abelian $U(1)_n$ charges may be carried by 
non-trivial matter representations of one or both ${\cal G}_{\rm obs}$ and
${\cal G}^{\rm NA}_{\rm hid}$ non-Abelian gauge groups.
Mixed states that transform under non-trivial representations of both
${\cal G}_{\rm obs}$ and ${\cal G}^{\rm NA}_{\rm hid}$ can be present. 

One method of string model construction relies on world-sheet free 
fermions
to represent the internal degrees of freedom necessary for conformal
anomaly cancellation \cite{ABKW,KLST}. In this construction, four dimensional 
heterotic\footnote{We choose the left-movers as the fermions 
assigned to carry world-sheet supersymmetry.} 
models of this type contain 18 internal left-moving real world-sheet 
free fermions, $\fer^{\re = 3,20}$ and
44 right-moving real world-sheet free fermions, ${\bfer}^{\re = 21,64}$,
in addition to the two left-moving real world-sheet fermions 
$\fer^{\re = 1,2}$ whose indices denote transverse spacetime directions.
 
In each free-fermionic model, a collection of boundary vectors 
$\{ \vec{\alpha}\}$ allowed by modular invariance
specifies the phase changes that the 64 world-sheet fermions $\fer^\re$ can 
undergo after they traverse the two non-contractible loops, 
$l_{\alpha}$ or $l_{\beta}$, on the genus-one world-sheet.
While there are two genus-one loops, there is
only one independent basis set of boundary vectors.
Modular invariance requires that the two sets of allowed boundary
conditions for the two loops are identical. 
Transporting a fermion $\fer$ around the loop $l_{\alpha}$ (or $l_{\beta}$) 
results in appearance of a phase  
\beq 
\fer^\re \rightarrow -\e^{i\pi\alpha_\re}\fer^\re, 
\label{ferp}  
\eeq
for rational $\alpha_\re$ in the range $-1< \alpha_\re\le 1$. 
For complex fermions 
$\fer^j_c \equiv \fer^{\re_1} + i \fer^{\re_2}$,\footnote{The 
indices $j$ and $\re$ denote complex and real fermions, respectively.}
these boundary vectors are associated with a charge lattice formed by
$\{\vec{Q}_{\vec{\alpha}}\}$,
where each charge vector has components 
\beq
(Q_{\vec{\alpha}})_j = \frac{\alpha_j}{2} + F_j.
\label{ferq}
\eeq
$F_j$ is a number operator for fermion oscillator excitations with eigenvalues
$\{0,\pm 1 \}$ for non-periodic fermions and $\{0,-1\}$ for periodic.

In heterotic strings (with the world-sheet supersymmetric sector 
as left-moving and the world-sheet bosonic sector as right-moving)
each right-moving complex fermion corresponds to a local $U(1)$ symmetry, 
whose massless generator is produced by the world-sheet simple current 
\beq
U_j= :{\fer}_c^{j*}{\fer}_c^{j}:\, .
\label{u1f}
\eeq  
Simple currents have normalizations 
\beq
\langle U_j , U_j \rangle = 1\, .
\label{u1fn}
\eeq 
In contrast, each left-moving complex fermion is only associated 
with a global $U(1)$ symmetry. 

Some of the 20 left-moving and 44 right-moving real world-sheet fermions
$\fer^{\re}$
cannot always be paired (as a result of differing boundary conditions)
to form left- or right-moving complex fermions. 
Instead, a left-mover and a right-mover may be paired to form a non-chiral
Ising fermion, or some left-movers and/or right movers may remain unpaired, 
forming chiral Ising fermions. In both of these cases, the  
boundary conditions are limited to $\alpha_{\re} = 0,1$.     
Clearly, models must contain an even
number of both left- and right-moving real Ising fermions. 
For every two right-moving Ising fermions
the rank of the gauge group is reduced by one
(independent of how these fermions divide into
chiral or non-chiral Ising types). 

To date, essentially only two primary  
embeddings of SM simple roots on charge lattices
have been used in free-fermionic models
(although many other embeddings are possible).
The SM embedding used in \cite{FNY1} and \cite{AF1}
(and all other ``NAHE'' class models \cite{NAHE})
is also the minimal embedding.
That is, the necessary root charges are obtained using
the lowest possible number of complex world-sheet fermions, which is 
five. No fewer than five are sufficient because
level-one $SU(n)$ roots can only be obtained by breaking
$SO(2n) \rightarrow SU(n)\times U(1)$ (the exception to this 
rule is associated with $SU(2)\times SU(2) \equiv SO(4)$).
Thus, the minimal free-fermionic SM embedding is, 
\beq
{\cal G}= SU(3)_C\times U(1)_C\times SU(2)_L\times U(1)_L\, ,
\eeq 
where the rank 3 algebra,
$SU(3)_C\times U(1)_C\in SO(6)$, originates in 
the charges of three complex fermions, $\bfer^{j= 1,2,3}_c$, 
and that of the 
rank 2 algebra, $SU(2)_L\times U(1)_L\in SO(4)$, 
from those of two additional complex fermions, $\bfer^{j= 4,5}_c$.
In terms of the five charges of the $3+2$ complex fermions,
the simple roots for $SU(3)_C$ and $SU(2)_L$ are
\beqn
SU(3)_C &:& (\phm 1,-1,\phm 0;\phm 0,\phm 0)
\label{su3rta1}\\
        & & (\phm 0,\phm 1, -1;\phm 0,\phm 0)
\label{su3rta2}\\
SU(2)_L &:& (\phm 0,\phm 0,\phm 0;\phm 1, -1). 
\label{su2rta1}
\eeqn

In contrast, the seven models in \cite{CHL} involve a
non-minimal SM charge embedding, requiring eight complex fermions,
\beqn
SU(3)_C &:& (\phm\half,\phm\half, -\half,-\half,-1,\phm 0,\phm 0,\phm 0)
\label{su3rtb1}\\
        & & (\phm    0,\phm    0,\phm  0,\phm 1,\phm\half,\phm\half ,-\half,\phm\half)
\label{su3rtb2}\\
SU(3)_C &:& (\phm\half,\phm\half, -\half,-\half,\phm 1,\phm 0,\phm 0,
\phm 0)\, .
\label{su2rtb1}
\eeqn
  
The general form of hypercharge candidates in minimal SM embedding models is  
\beq
U(1)_Y = b_C U(1)_C + b_L U(1)_L  + \sum_n b_n U(1)_n\, , 
\label{hyperQm}
\eeq
where the $b$'s are rational coefficients.
It has been argued by claims of phenomenological necessity\cite{AFT}
that viable hypercharge definitions are limited to
\beq
b_C= \frac{1}{3};\,\, b_L= \pm \frac{1}{2};\,\, b_n= 0.
\eeq
Generally the
``$+$'' sign is chosen, as is true in particular for 
\cite{FNY1} and \cite{AF1}. The choice of the opposite sign corresponds to
an exchange of particle identities $\uc\lra\dc$, $\ec\lra\vc$,
and $H_U\lra H_D$.

In the non-minimal SM embedding of \cite{CHL}  
there are no apparent special Abelian symmetries like $U(1)_C$ and $U(1)_L$,
so the generic expression for a hypercharge candidate is simply
\beq
U(1)_Y = \sum_n b_n U(1)_n\, . 
\label{hyperQm2}
\eeq  
Phenomenologically viable possibilities for $b_n$ in this class will be 
analyzed in the subsequent section, including those presented for models CHL4 through CHL6 in \cite{CHL}.

Generally the currents $U_n$ of the Abelian symmetries $U(1)_n$ 
are not themselves simple currents $U_{j}$ 
as defined in (\ref{u1f}) and (\ref{u1fn}).
Instead, most of the $U_n$ are linear combinations of simple currents, 
\beq
U_n = \sum_j b'_{n,j} U_j\, .
\label{unsc}
\eeq      
Often such linear combinations are not normalized 
to one, {i.e.}, 
\beq
\langle U_n,\ U_n\rangle \neq 1\, .
\label{unscn}
\eeq

Differing hypercharge definitions can    
significantly alter the phenomenology of the effective field theory. 
In particular, the definition determines the value of $k_Y$ for 
$U(1)_Y$ \cite{knorm}.\footnote{In string theory $k_Y$  
relates the definition of hypercharge $Y$ to the hypercharge 
contribution $h_Y$ to the total conformal dimension, 
$ h_Y = \frac{Y^2/k_Y}{2}$ ,  
of a physical state.}  
From the low-energy point of view, the value of $k_Y$ is especially
important with regard to shifts in the 
string unification scale $\MS$.
In string models, $k_Y$ and the corresponding levels, $k_3$ and $k_2$, of 
$SU(3)_C$ and $SU(2)_L$, are related at tree level by
\beq
   g^2_i k_i = g^2 = g^2_{\rm string}/2,
\label{gtree}
\eeq
where $g_{Y,2,3}$ are
the canonical gauge couplings of $U(1)_Y$, $SU(2)_L$, and $SU(3)_C$,
respectively defined so that $Y= \frac{1}{6}$ for a quark doublet $Q_L$
and $\tr\, t_a t_b = \frac{1}{2}\delta_{ab}$ (corresponding to 
the highest root $\psi$ normalized to $\psi^2= 1$) 
for the generators $t_a$ of the 
fundamental representations of $SU(2)$ and $SU(3)$.
At one-loop, these gauge couplings obey the renormalization group
equations of the effective field theory: 
\beq
\frac{16\pi^2}{g_i^2(\mu)} = k_i\frac{16\pi^2}{g^2}
                           + b_i\ln\frac{M^2_{\rm string}}{\mu^2} + \Delta_i,
\eeq
for $g_i= g_Y,\, g_2,\, g_3$.
$b_i$ are the one-loop beta-function coefficients and $\Delta_i$ are
``threshold'' corrections from the infinite tower of massive string states.
$\MS \sim 5\times 10^{17}$ GeV is the one-loop corrected coupling
unification
scale \cite{Kap}.

$k_Y$ can be determined by studying a fermion-fermion-gauge coupling.
While $k_i$ is limited to positive integer values for NAB groups, 
$k_i$ is not necessarily integer in the Abelian case, 
but can take on positive rational values. 
When $SU(3)_C\times SU(2)_L$ is embedded in $SU(5)$ (or is
embedded in $SU(5)$ in a string model,
but $SU(5)$ is broken by string boundary conditions -- as in
minimally embedded ``NAHE'' free-fermionic models \cite{NAHE}),
the lowest possible value for $k_Y$ is $\frac{5}{3}$.
Note however that values of $k_Y < \frac{5}{3}$ lead to somewhat better 
agreement between the string scale and MSSM unification scale (assuming
the MSSM particle content).

For a generic hypercharge 
\beq
Y \equiv  \sum_n b_n Q_n  =  \sum_j b_j Q_j,
\label{hypdef1}
\eeq
$k_Y$ can be expressed as \cite{CHL,DF},
\beq
% k_Y = {2}(\frac{1}{6 Y(Q)})^2\sum_n b_n^2 \langle U_n , U_n \rangle, 
  k_Y = {2}\sum_n b_n^2 \langle U_n , U_n \rangle 
      = 2\sum_j b_j^2\, ,
\label{k1def2}
\eeq
where each $U_n$ is the current associated with the corresponding 
$U(1)_n$, which are linear sums of the $U(1)_j$ with 
unit norm. The factor of two
in eq.\ (\ref{k1def2}) is a result of differing string and 
field-theoretic conventions with regard to traces of non-Abelian
gauge generators \cite{Kap} (and the related highest root normalizations). 
While the canonical field-theoretic choice
is $\tr\, t_a t_b = \frac{1}{2}\delta_{ab}$ (with $\psi^2= 1$) 
for $SU(n)$ gauge groups, in string theory the corresponding choice   
is $\tr\, t_a t_b = \delta_{ab}$ (with $\psi^2= 2$) 
instead\footnote{A factor of two was not included in the definition of 
hypercharge level used in \cite{CHL}. 
Thus, the correct values of the $k_Y$'s in the models presented in \cite{CHL}
are actually twice that given in \cite{CHL}, and are all 
greater than $\frac{5}{3}$.}.
When the gauge group includes an anomalous $U(1)$ there is a simpler way
of computing $k_Y$, as will be explained in the next sub-section.
 
%The expression for the effective level simplifies when the hypercharge 
%is defined directly in terms of the charges 
%$Q_j$ defined in the eq.\ (\ref{ferq}), which  correspond to
%the (up to) 22 right-moving simple currents $U_{j\ge 21}$.
%Since these simple currents have unit norms, 
%$\langle U_{j},\,  U_{j}\rangle = 1$, 
%(\ref{k1def2}) reduces to
%\beq
% k_Y  
%\label{k1def3}
%\eeq
%for
%\beq
%\label{hypdef2}
%\eeq

\subsection{Anomalous $U(1)$ and Flat Directions}

The appearance of anomalous $U(1)$'s in four-dimensional
string models has been discussed extensively 
\cite{KN,CF,CCEEL2,genanoms,genanomp}. 
In the original free-fermionic charge basis of (\ref{ferq}), 
models usually contain more than one $U(1)_n$ with $\Tr Q_n\neq 0$.
However, the anomaly can be transferred into a single
$U(1)_A$ through the unique rotation  
\beq
         U(1)_{\rm A} \equiv c_A\sum_n \{\Tr Q_{n}\}U(1)_n,
\label{rotau1}
\eeq
with $c_A$ a normalization factor.
Following this rotation, a complete orthogonal basis 
$\{U(1)_a\}$ may be formed  
from the non-anomalous components of the original set of $\{U(1)_n\}$.

The elimination of all triangle anomalies except those involving
one or three $U(1)_A$ gauge bosons is guaranteed by  
the Green-Schwarz (GS) relations,  
\beqn
\frac{1}{k_mk_A^{1/2}}\mathop{\Tr}_{G_m}\, 
T(R)Q_A &=& \frac{1}{3k_A^{3/2}}\Tr Q_A^3
                      = \frac{1}{k_a k_A^{1/2}}\Tr Q_a^2Q_A 
                      =\frac{1}{24k_A^{1/2}}\Tr Q_A
                           \equiv 8\pi^2 \delta_{\rm GS} \, ,
\label{gsa}\\
\frac{1}{k_mk_a^{1/2}}\mathop{\Tr}_{G_m}\, 
T(R)Q_a &=& \frac{1}{3k_a^{3/2}}\Tr Q_a^3
                     = \frac{1}{k_Ak_a^{1/2}}\Tr Q_A^2 Q_a 
                     = \frac{1}{(k_a k_b k_A)^{1/2}}\Tr Q_a Q_{b\ne a} Q_A
\nolabel\\ 
        &=&\frac{1}{24k_a^{1/2}}\Tr Q_a = 0 \, ,
\label{gsna}
\eeqn
%and additional generalizations involving 
%$Q_A Q_a Q_b$, $Q_a Q_b Q_c$, etc..
where $k_m$ is the level of the gauge group $G_m$ and
$2 T(R)$ is the index of the representation $R$, defined by
\beq
\Tr\, T^{(R)}_a T^{(R)}_b = T(R) \delta_{ab}\, .
\label{tin}
\eeq
In a generic field-theoretic model, (\ref{rotau1}) would not
necessarily place the entire anomaly into a single $U(1)_A$.
The GS relations result from stringy modular invariance constraints
and guarantee the consistency of the model.
%The GS relations also may not hold in strongly coupled string models.
The physical content behind these relations is that the mixed anomalies
are canceled by the pseudoscalar partner of the string dilaton, which
couples universally to all gauge groups.
The relations (\ref{gsa}) can be used to compute $k_Y$ once the
states in the massless spectrum and their charges are known, without
further knowledge of the string origin of each state \cite{luis}.

The standard anomaly cancellation mechanism
\cite{DSW,ADS,DKI,AS} generates a FI  $D$-term,
\beq
%      {\e}^\phi \MP^2 \delta_{\rm GS}=
\xi = \frac{g^2_{\rm string}\Mp^2}{192\pi^2}\Tr Q_A\, ,
\label{fid}
\eeq
where 
%$\phi$ is the dilaton, $ g\equiv {\e}^{\phi/2}$ 
%is the physical four-dimensional string 
% %gauge% coupling, and
$\Mp = M_{\rm Planck}/\sqrt{8\pi}$ with
$M_{\rm Planck} \sim 1.2\times 10^{19}$ GeV.

The FI $D$- term is calculable in perturbative string theory,
since it is a genus-one string effect when determining masses \cite{ADS}
(and a genus-two effect when calculating the dilaton tadpole \cite{AS}).
The FI $D$- term triggers a shift to a nearby deeper vacuum with
non-zero VEVs for the scalar components $\varphi_i$ of supermultiplets
$\Phi_i$ such that the $D$-flatness constraints are
satisfied\footnote{Our convention
for defining $D_{\rm A}$ is that the corresponding $D$- term in the
Lagrangian is $\frac{1}{2k_A}g^2D_{\rm A}^2$, and similarly for $D_a$.}
\beq
D_{\rm A} = \sum_i Q^{(A)}_i |\varphi_{i}|^2 + \xi
%+ \frac{g^2 \Mp^2}{192\pi^2}{\rm Tr}Q_{\rm A}
= 0\,\, ,
\label{anomd}
\eeq
\beq
D_{a} = \sum_{i} Q^{(a)}_{i}|\varphi_{i}|^2 = 0,
\label{dif}
\eeq
along with $F$- flatness,  
\beq
F_{i} = \frac{\partial W}{\partial \Phi_{i}} = 0; \,\, W  =0.
\label{ff}
\eeq

\subsection{String Selection Rules for Superpotential Terms}

The (perturbative) superpotential for the low-energy effective field
theory 
of an underlying string model is significantly more constrained than
a generic field-theoretic superpotential, 
resulting from additional world-sheet symmetries 
than simply those that translate into standard gauged spacetime symmetries. 
Consequently, stringy superpotentials will generically have fewer
terms at a given order than one would expect simply from gauge group
charge conservation.  

Coupling coefficients for a superpotential term of order 
$(K+3)$, with $K\geq 0$, can be cast in terms of a 
$(K+3)$--point tree-level string amplitude $A_{K+3}$ of the 
form,\cite{KLNF,CCEEL1}
\beq
%A_{K+3} =  \left(\frac{g}{2\pi}\right)^{K}\frac{ g\eta C_K I_K}{\MP^{K}} \, ,
 A_{K+3} =  g \frac{C_K I_K}{(\pi \Mp)^{K}} \, ,
\label{ampk3a}
\eeq 
where $g$ is the gauge coupling at $\MS$ and $C_K$ is a coefficient of
${\cal O} (1)$ that encompasses different renormalization factors in the 
OPE of string vertex operators (including the
target space gauge group Clebsch-Gordan coefficients).
For the renormalizable $(K=0)$ trilinear terms, the $I_0$ factor 
reduces to a correlation function $f_{0}$ of world-sheet coordinates,
whereas for non-renormalizable $(K>0)$ terms, $I_{K}$ is an integral of 
a correlation function 
$f_{K}$ over $K-3$ world-sheet coordinates
\beq
I_K=\int d^2z_3\cdots d^2z_{K+2} 
f_K(z_1=\infty,z_2=1,z_3,\cdots,z_{K+2},z_{K+3}=0),
\label{igen}
\eeq
where $z_i$ is the world-sheet coordinate of the  
conformal dimension $(1,1)$ vertex operator $V_i$  
for the $i^{\rm th}$ physical heterotic string state. 
By $SL(2,C)$ world-sheet invariance, three coordinates may be fixed. 
Generally, $z_1\rar \infty$, $z_2 = 1$, and $z_{K+3} = 0$ is chosen.   

In the free-fermionic construction,
the vertex operator $V_i$ may be factored into a product
of vertex operators associated with  
(i)  the Lorentz spacetime momentum factor, $V_i^{mom}$;
(ii) the Lorentz spacetime spin factor, $V_i^{spin}$;
(iii) the BRST superconformal ghost charge, $V_i^{BRST}$;
(iv) the global left-moving and local right-moving Abelian symmetry groups, 
$\prod_{n'} V_i^{U(1)^{global}_{n'}}$ and $\prod_n V_i^{U(1)^{local}_n}$,
respectively; 
(v) the local right-moving NAB symmetry groups 
$\prod_m V_i^{{\cal G}_m}$;
(vi) the non-chiral Ising functions, $\prod_{q}V_i^{NCI_q}$; and 
(vii) the chiral Ising functions $\prod_{q'} V_i^{CI_{q'}}$.

The correlation function $f_K$ similarly factors into products
of correlation functions for each of these classes of vertex 
operators \cite{KLNF},
%the $K+3$ fields.
\beqn
f_K(z) &=& \lan \prod_{i=1}^{K+3}V_i^{mom}\ran
           \lan \prod_{i=1}^{K+3}V_i^{spin}\ran
           \lan \prod_{i=1}^{K+3}V_i^{BRST}\ran
\prod_{n'} \lan \prod_{i=1}^{K+3}V_i^{U(1)^{global}_{n'}}\ran
\prod_{n}  \lan \prod_{i=1}^{K+3}V_i^{U(1)^{local}_{n}}\ran
\nolabel\\
       & &
\prod_{m}  \lan \prod_{i=1}^{K+3}V_i^{{\cal G}_m}\ran
\prod_{q}  \lan \prod_{i=1}^{K+3}V_i^{NCI_q}\ran
\prod_{q'} \lan \prod_{i=1}^{K+3}V_i^{CI_{q'}}\ran\, \, .
\label{fkcor}
\eeqn

The spacetime spin correlator for two spacetime fermions and $K+1$ scalars
is trivial, contributing only a factor of 
\beq
\langle S_{\alpha}(z_1) S_{\beta}(z_2) \rangle = (z_{1}-z_2)^{-1/2} 
\label{spincor}
\eeq 
to $f_K$,
where $S_\alpha$ is a conformal field representing a Lorentz spinor.
With the exception of the Ising correlation functions,
the remaining correlation functions in the vertex operators
have exponential form. 
For an Abelian symmetry or the BRST ghost charge it is 
\beq
\langle
      \prod_{i} {\rm e}^{i Q_i H}
\rangle
=     \prod_{i<j} z_{ij}^{Q_i Q_j}\, ,  
\label{expcoragh}
\eeq
while for a non-Abelian symmetry the correlation function is
\beq
\langle
      \prod_{i} {\rm e}^{i \vec Q_i\cdot \vec J}
\rangle
=     \prod_{i<j} z_{ij}^{\vec Q_i \cdot \vec Q_j}\, ,  
\label{expcorna}
\eeq
where $z_{ij} \equiv z_i - z_j$ (in this language, $Q_i= -ic$ is imaginary
for ghost systems).
Non-conservation of any (local or global) Abelian 
or non-Abelian charge, i.e., a case of 
$\sum_{i} Q_i \neq 0$ or $\sum_{i}\vec{Q}_i\neq\vec{0}$,
yields $I_K=0$. 
On the other hand, the vertex operators must contribute
a total BRST superconformal ghost charge of $-2$ to 
cancel the superconformal ghost charge carried by the vacuum \cite{LT}.
The spacetime momentum correlation function is
\beq
\langle
      \prod_{i} {\rm e}^{i \half \vec K_i\cdot \vec X_i}
                {\rm e}^{i \half \vec K_i\cdot \vec {\bar X}_i}
\rangle
=     \prod_{i<j} | z_{ij} |^{\half \vec{K}_i \cdot \vec{K}_j}\, .  
\label{expcorstk}
\eeq

Ising correlators (both non-chiral and chiral classes) are non-trivial. 
For example, there are six types
of conformal fields (including the identity operator) 
associated with a non-chiral 
Ising fermion: a left-moving real world-sheet fermion $f(z)$, 
its right-moving counterpart $\ov{f}(\ov{z})$, 
the energy operator $\epsilon (z,\ov{z})\equiv f\ov{f}$, and
spin fields $\sigma_{+}(z,\bar z)$ and $\sigma_{-}(z,\bar z)$
(also known as order/disorder operators).
Correlators involving the spin fields of a given Ising fermion are non-zero 
if and only if they can be factored into combinations of 
\beq
\lan\sigma_{+}\sigma_{+}\ran,\quad  \lan\sigma_{-}\sigma_{-}\ran,\quad  
\lan\sigma_{+}\sigma_{-}f\ran,\quad {\rm and}\quad 
\lan\sigma_{+}\sigma_{-}\ov{f}\ran,  
\label{cising}     
\eeq
while
correlators not involving the spin fields require an even number of both
$f$ and $\ov{f}$.

Chiral Ising correlation functions have additional subtleties over non-chiral 
Ising correlators, but under certain conditions\cite{KLST,CHLI}, they may be 
represented in terms of vertex operators of ``broken $U(1)$ charges.'' 
When these conditions are satisfied (such as in the CHL models), the chiral 
Ising fermions may actually be paired in a vertex operator and associated
with a ``broken charge'' pair $\pm |Q|$. Then the vertex operator 
of the set of all chiral Ising fields, can be written for a physical state as 
\beq
V_i^{CI}= \prod_{q'} V_i^{CI_{q'}},
\eeq
whose two
charge vectors, $\vec{Q}^{(i)}$ and $-\vec{Q}^{(i)}$,
differ only by an overall sign. 
The correlation function $\lan \prod_{i=1}^{K+3} V_i^{CI} \ran$ 
is nonzero when there is a choice of signs such that
$\sum_i \pm \vec{|Q|}^{(i)}= \vec{0}$.  

Both conservation of global world-sheet charges
and the Ising field correlation selection rules are truly 
stringy effects.
%, without parallels in standard field theory.
In Section IV, we shall see examples in which these stringy effects 
reduce the number of 
superpotential terms at a given order otherwise allowed 
by gauge invariance \cite{GC}.   

\subsubsection{Picture Changing and Charge Conservation}

Satisfying both conservation of global world-sheet charges 
and the Ising field selection rules of (\ref{cising}) 
in the superpotential terms of the effective field theory of a 
string model has its subtleties.
The complications stem from cancelling the vacuum
BRST ghost charge\cite{GSW,LT} (mentioned above).

The correlation function $f_K$ is generated by the product of the
vertex operators $V_i$ for the $K+3$ superfields $\Phi_{i}$ forming the
candidate superpotential term,
\beq
f_K = \lan V^f_1(z_1\rar \infty) V^f_2(z_2=1) 
      V^b_3(z_3) \cdots V^b_{K+3}(z_{K+3}=0)\ran \, ,
\label{fkv}  
\eeq
where $V^{f\,(b)}_i$ is the fermionic (bosonic) part of the complete
superfield vertex operator $V_i$. 
The BRST ghost charge associated 
with the canonical fermionic (bosonic) vertex operator is $-\half$ ($-1$).
Conformal invariance allows a physical vertex operator with 
a given ghost charge
to be ``picture-changed'' into an equivalent vertex operator with a new  
ghost charge differing by an integer value from the first. 
However, as we discussed, cancellation of the vacuum ghost charge anomaly  
requires that the net ghost charge for $f_K$ be $-2$.
Since in the canonical picture the net ghost charge is $-2+(3-K)$,
the last $K$ bosonic vertex operators should be picture-changed to
carry a ghost charge of 0.\footnote{Any other set of picture-changes that
similarly yield a net ghost charge of $-2$ for $f_K$ would also be acceptable
and would generate the same superpotential terms.}
That is,
\beq
       V^b_{4\, (-1)} \cdots V^b_{K+3\, (-1)}\rar
       V^b_{4\, (0)} \cdots V^b_{K+3\, (0)}\, ,
\label{pcv}  
\eeq
with ghost charge of a vertex operator 
explicitly denoted by the subscript in parenthesis.

The spin-$\frac{3}{2}$ supercurrent $T_{3/2}$
of the $N=2$ world-sheet supersymmetry
acts as the picture-changing operator for a superfield vertex operator, 
increasing the vertex operator ghost charge 
by one unit:
\beq
V_{(c+1)}(z)= \lim_{w\rar z} \exp^{c}\, T_{3/2}(w) V_{(c)}(z)\, .
\label{pco}   
\eeq
$T_{3/2}$ can be separated into three components distinguished by their
respective charges (appearing as superscripts)
under the $U(1)_{N=2}$ current (of the $N=2$ global world-sheet supersymmetry) 
also present in the $N=2$ algebra: 
\beq
T_{3/2} = T^0_{3/2} + T^{-1}_{3/2} + T^{+1}_{3/2}\, , 
\label{tft}
\eeq
where the superscripts denote world-sheet charges. 
The canonical bosonic (fermionic) vertex operators 
with ghost charge $-1$ $(-\half)$ also carry
$+1$ $(-\half)$ charge under $U(1)_{N=2}$. 
Thus, when acting on the ghost charge $-1$ vertex operators, 
 only the $T^{-1}_{3/2}$ component of $T_{3/2}$  
will lead to conservation of $f_K$'s total $U(1)_{N=2}$ charge in 
(\ref{fkcor}) and (\ref{fkv}). 
 
When all 20 left-moving real world-sheet fermions $\fer^{j= 1,20}$ have only
periodic/antiperiodic boundary conditions,\footnote{$N=2$ algebras for 
free-fermionic models containing left-moving fermions with 
rational, non-integer boundary conditions have been studied in \cite{GCS}.}
the 18 internal fermions can be regrouped into six sets of three fermions 
(and appropriately relabelled):
\beq
\{ (\fer^1\fer^2),\,\,(x,y,\w)^{i=1\,\, {\rm to}\,\, 6} \}\, . 
\eeq  
While
spacetime supersymmetry requires fermions $x^{2j-1}$ and $x^{2j}$ to
combine into complex fermions 
$X^{2j-1,2j} \equiv \isqrt (x^{2j-1} +i x^{2j})$, for $j= 1$ to $3$, 
the various $y^i$ and $\w^i$ are unconstrained and 
may form either complex or Ising types. 
In terms of the 20 real fermions  
(and two transverse world-sheet bosons $X_{\mu=1,2}$)
the world-sheet supercurrent is
\beq
T_{3/2}(z) = \fer^{\mu}\partial X_{\mu} 
          + i\sum_{i=1}^6 x^i y^i \w^i
\label{t32a}
\eeq
while
the $U(1)_{N=2}$ generator (connected with spacetime supersymmetry) is 
\beq
J(z) = :X^{12*}X^{12}: + :X^{34*}X^{34}:  + :X^{56*}X^{56}:\, .   
\label{j2a}
\eeq
Eq.\ (\ref{j2a}) indicates
that the global $U(1)_{N=2}$ charge of a state is the sum of all 
three of its $X^{2j-1,2j}$ charges.
Complex fermions like the $X$ can be replaced by real bosons $S$, 
using the $U(1)$ current boson/fermion identity, 
\beq
i\partial_z S(z) = :X^* X:\, .
\label{bfe}
\eeq
This equivalence allows $J(z)$ to be expressed as
\beq
 J(z) = i\partial_z (S_{12} + S_{34} + S_{56})\, . 
\label{n2j}
\eeq

A related bosonized form of $T_{3/2}(z)$ is easily separated into 
the three components identified in (\ref{tft}) by their $U(1)_{N=2}$ charges:
\beqn
T^0_{3/2}(z) &=& \fer^{\mu}\partial X_{\mu}
\label{t32u1}\\
T^{-1}_{3/2}(z) &=&        \exp^{-i S_{12}}\tau_{12}
                          +\exp^{-i S_{34}}\tau_{34}
                          +\exp^{-i S_{56}}\tau_{56}
\label{t32u2}\\
T^{+1}_{3/2}(z) &=& -(T^{-1}_{3/2}(z))^{*}, 
\label{t32u3}
\eeqn
where,
\beq
\tau_{ij}\equiv \frac{i}{\sqrt{2}}(y^i \w^i + iy^j \w^j). 
\label{taueq}
\eeq 

The form of $T^{-1}_{3/2}$ in (\ref{t32u2}) implies 
that picture-changing 
a $V^b_{-1}$ operator into a $V^b_{0}$ operator via (\ref{pco}),    
$T^{-1}_{3/2}$ simultaneously alters
both the $U(1)_{N=2}$ charge and some $y$- and $\w$-related 
charges or Ising fields. 
Consider, for example, the effect of the first component, 
$\exp^{-i S_{12}}(y^i \w^i)$, of $T^{-1}_{3/2}$ on a generic
$V^b_{-1}$. 
The operator $\exp^{-i S_{12}}$ decreases the $X^{12}$ charge  
(and thus also the total $U(1)_{N=2}$ charge) by 1. 
Further, if $y^1$ is an Ising fermion, then a spin field 
$\sigma^{y^1}_{+}$ ($\sigma^{y^1}_{-}$) in $V^b_{-1}$ will be 
converted into the opposite spin field type, 
$\sigma^{y^1}_{-}$ ($\sigma^{y^1}_{+}$), in $V^b_{0}$
by the $y^1$ factor in this $T^{-1}_{3/2}$ component.
If, however, there were no such Ising spinor in $V^b_{0}$, 
then a new Ising fermion excitation $y^1$ would appear in $V^b_{0}$.   
On the other hand, if $y^1$ was part of a complex fermion, e.g., 
$Y^{1,3}\equiv \isqrt (y^1 + i y^3)$,
then the $y^1$ operator in the $T^{-1}_{3/2}$ component would
create two separate terms in $V^b_{0}$: 
one term would have its $Y^{1,3}$ charge raised by one unit and 
the other term would have its $Y^{1,3}$ charge lowered by one unit. 
Raising and lowering of the charge would both occur
because in terms of $Y^{1,3(*)}$,    
$y^1 = \isqrt (Y^{1,3} + i Y^{1,3*})$. That is, $y^1$ 
contains both charge raising and lowering operators.
The $\w^1$ in factor $T^{-1}_{3/2}$ component 
would act similarly on $V^b_{-1}$. 

%As emphasized above, conservation
%of left-moving global $U(1)$ charges and Ising 
%correlation functions for candidate non-renormalizable superpotential terms 
%must be tested {\it after} picture-changing rather than before.
Therefore, 
terms that seem to be allowed (disallowed) prior to picture-changing may 
actually be disallowed (allowed). Of particular importance is that
picture-changed bosonic vertex operators $V^b_{0}$
can contain several different terms due to  
(i) the six separate terms in $T^{-1}_{3/2}(z)$ and 
(ii) operators in each term of $T^{-1}_{3/2}(z)$
that can sometimes act as both 
raising and lowering operators of $y$- and $\w$-associated charges.  
When a correlation function $f_K$ is under examination, 
all possible combinations of terms in the
$K$ pictures-changed vertex operators must be considered.

%For a gauge invariant product of $K>3$ superfields actually to 
%appear in the stringy superpotential,
%global charge conservation and Ising selection rules need be 
%simultaneously satisfied in (\ref{fkv}) only  
%for a single product combination of picture-changed vertex operator terms. 
%We will show by specific example, how picture changing affects which 
%gauge-invariant combination of superfields can appear in a 
%stringy superpotential. 
We now illustrate the technique explicitly, by considering a specific example.
The free-fermionic model presented in
\cite{FNY1} contains among its $N=1$ spacetime superfields four denoted as 
$H_{39}$, $H_{37}$, $H_{32}$, and $H_{30}$ 
(these fields are identified in our Tables~IIIa and ~IIIb as
$S_{43}$, $S_{28}$, $S_{24}$, and $S_{7}$, respectively). 
Their gauge charges can be found in 
 Table 2 of \cite{FNY1}.
This is the same set of four non-Abelian singlets
for which in \cite{CCEEL1} we computed the integral $I_1$ of the correlation
function $f_1$. In Table~I we list left-moving global $U(1)$ charges
and non-chiral Ising fields in the fermion components of the superfields
$H_{39}$ and $H_{37}$ and in the bosonic components of 
$H_{32}$ and $H_{30}$.

The model in \cite{FNY1} contains six complex left-moving world-sheet fermions: 
$\fer^{12}= \fer^1 + i\fer^2$,  
$   X^{12}= x^1 + i x^2$,  
$   Y^{16}= y^1 + i\w^6$,
$   W^{13}= y^1 + i\w^3$,
$   X^{34}= x^3 + i x^4$, and  
$   Y^{36}= y^3 + i y^6$. 
Table~I only lists charges under the first
real fermion component of a complex fermion. 

Additionally, the model possesses six non-chiral Ising pairs:
$( y^2,\bfer^{38})$,
$(\w^2,\bfer^{44})$,
$( y^4,\bfer^{40})$,
$(\w^4,\bfer^{46})$,
$( y^5,\bfer^{41})$, and
$(\w^5,\bfer^{47})$.
There are no
anti-periodic excitations from any of the right-moving $\bfer$ components in
non-chiral Ising fermion pairs for any of the four superfields. Thus, the
corresponding $\bfer$ are not relevant to the picture-changing discussion
below. 

%Since there are four fields, we must picture-change exactly one 
%field. Which one we choose is unimportant: if none of the terms
%in $T^{-1}_{3/2}$ can result in all non-zero
%correlations functions within $f_1$ when picture changing the chosen field, 
%then $T^{-1}_{3/2}$ acting on any of the three other fields can do 
%no better. Thus, 
We choose to picture change $H_{30}$. From Table~I
we see that the net $S_{34}$ and $S_{56}$ charges are both zero before
picture changing, while the net $S_{12}$ charge is $+1$. Thus,
we must use the $\exp^{-i S_{12}}\tau_{12}$
component of $T^{-1}_{3/2}$ to picture change $H_{30}$ and cancel the
$S_{12}$ pre-picture changed charge.  
Before picture changing,
the Ising field correlations for $( y^2,\bfer^{38})$ and $(\w^2,\bfer^{44})$ 
are non-zero ($\langle \sp \sp \rangle$ and $\langle \sm \sm \rangle$,
respectively), while the net $Y^{16}$ and $W^{13}$ 
global $U(1)$ charges are both $-1$ and $-1$, respectively. 

The $y^2$ and $\w^2$ Ising correlations 
remain unaffected by the $y^1 \w^1$ component of 
$\tau_{12}\equiv \frac{i}{\sqrt{2}}(y^1 \w^1 + iy^2 \w^2)$. 
In contrast, the raising operator in $y^1$ cancels the 
pre-picture change net $Y^{16}$ charge by altering
$H_{30}$'s $Y^{16}$ charge from $\mh$ to $\ph$.  
The raising operator in $\w^1$ similarly cancels the $W^{13}$ charge.
%Thus, while prior to picture changing these four
%superfields would not have appeared as a superpotential term,
%they do appear after picture changing.
 
%We point out that the same picture changing operator,
%$\exp^{-i S_{12}}\frac{i}{\sqrt{2}}(y^1 \w^1)$ could instead have acted on
%any of the other fields to produce similar results. Acting on the fermionic 
%fields of either  
%$H_{39}$ or $H_{37}$ would have converted a canonical
%$V^f_{(\mh )}$ vertex operator into a non-canonical  
%$V^f_{(\ph )}$ vertex operator. 

Determining if the string amplitude is non-zero is
straightforward for low values of $K$.
This process might seem unwieldy for increasingly higher values 
of $K$; however, simple arguments based on invariance \cite{LT}
of outcome under differing choices of picture-changed fields 
permit a more tenable approach which does not require that 
picture changing be performed on $(K-3)$ individual states.  
Instead, essentially only the total picture-changing effect from 
$K$ various components of $T^{-1}_{3/2}$ need act on an effective 
``composite field.'' 
The charges and Ising fields of the effective composite state are
the summations of the separate respective charges of the 
original $K+3$ {\it canonical} fields. 
The ``composite field'' charges for the $W_4$ example above 
are given in the ``net charges'' row of Table I. A zero net 
``charge'' appears in an Ising field column if the Ising correlation 
function of the four fields is already non-zero.

For a non-zero string amplitude, there must be an allowed set of 
$K+3 = \sum_{i=1}^6 P_{i}$, for $P_i\in \{{\rm Z}^+, 0\}$, 
picture changing operators
\beqn
\sum_{P_{1}} \exp^{-i S_{12}}(y_1 \w_1) &+&
i\sum_{P_{2}} \exp^{-i S_{12}}(y_2 \w_2) +
\sum_{P_{3}} \exp^{-i S_{34}}(y_3 \w_3) +
i\sum_{P_{4}} \exp^{-i S_{34}}(y_4 \w_4) +
\nolabel\\
\sum_{P_{5}} \exp^{-i S_{56}}(y_5 \w_5) &+&
i\sum_{P_{6}} \exp^{-i S_{56}}(y_6 \w_6) ,
\label{netpca}
\eeqn
that both contains an appropriate set of total $U(1)$ charges to cancel 
the corresponding charges of the ``composite field'' and 
provides for non-zero Ising $y$'s and $\w$'s correlations. 

Cancellation of the total $U(1)_{N=1}$ charge from two canonical fermion fields
and $K-2$ canonical scalar fields, 
$Q^{net}_{N=2}= 2\times (-1/2) + (K-2)\times (1) = K-3$,
is trivially accomplished by any $\{ P_i\}$ set
specifying the picture-changing operators in (\ref{netpca}), since
$(\sum_{i=1}^6 P_i )\times (-1) = - (K+3)$.
However, separate cancellation of the
three charge components $Q^{net}_{i,i+1}$ of $Q^{net}_{N=2}$ 
associated with the three $S_{i,i+1}$,  
imposes the first requirement for a good picture-changing charge cancellation:
each $Q^{net}_{i,i+1}$ must be a positive integer. 
Then cancellation of each $Q^{net}_{i,i+1}$ requires 
choices of $P_i$ and $P_{i+1}$ such that
\beq
(P_i + P_{i+1}) (-1) = - Q^{net}_{i,i+1}.     
\label{pic}
\eeq  

Values of the six $P_i$ 
must lead to cancellation of all $y_i$/$\w_i$-related charges.
Our previous discussion of $\tau_{i,i+1}$ operators 
implies that $P_i$ and $P_{i'}$ lead to
cancellation of a particular charge $Q_{i,i'}$ associated with
a complex fermion formed from real fermions $f_i$ and $f_{i'}$,
(where $f$ is  $y$ or $\w$) if and only if
\beqn
Q_{i,i'} &=&  P_i + P_i'\quad ({\rm mod}\,\, 2)      
\label{qpp}\\  
| Q_{i,i'}| &\leq&  P_i + P_i'.
\label{abqpp}
\eeqn
Searching for possible $\{P_i\}$ solutions to these constraints
and corresponding Ising-related constraints can be easily 
performed by a simple computer subroutine, making an efficient
determination of high order terms in the superpotential feasible.

\section{Flat Direction Analysis}

\subsection{Classification of Fields}

Our strategy to find the set of flat directions which satisfy the flatness
constraints (\ref{anomd}), (\ref{dif}), and (\ref{ff})
is the following.  First, the moduli space
of flat directions under the nonanomalous $U(1)$'s only is determined.
This space of flat directions can be described by a basis of independent
holomorphic gauge invariant (under the non-anomalous $U(1)$'s) monomials
(HIM's) \cite{LTGSM}, 
or equivalently by a larger
superbasis\footnote{The elements of this superbasis are not linearly
independent.  However, it has the advantage that any flat direction can be
expressed simply as a product of the elements in the superbasis.} 
of all one dimensional HIM's (i.e., with one free VEV unconstrained by the 
flatness conditions).
The HIM's of the basis and the superbasis are then classified according to
their anomalous charge. If the sign of the anomalous charge of a given
HIM is opposite to that
of the FI term some free VEV in the flat direction will adjust itself
to cancel the anomalous $D$-term ($D_A$).
The $F$-flatness conditions are then addressed for this $D$- flat
direction.

However, for the purpose of determining if there exist NAB singlet flat
directions which can cancel $D_A$, there is a useful
classification of the fields in the model which can show immediately if
such a flat direction is possible, as we explained in \cite{CCEEL2}.  For
the sake of completeness, we repeat the strategy of this classification
here. One defines an auxiliary charge $\overline{Q}$ as a linear
combination of non-anomalous $U(1)$ charges 
\be
\overline{Q}_j\equiv\sum_{a=2}^m\alpha_aQ^{(a)}_j,
\ee
where the $\alpha_a$ are chosen for convenience trying to maximize the
number of fields for which $Q_A=\overline{Q}$.
This relation cannot hold for all the chiral fields in the model
so we define
the quantities
\be
\hat Q_j=Q_j^{A}-\overline{Q}_j,
\ee
and classify all the chiral fields in three
different types, depending on the sign of $\hat Q_j$ :
\bear
&&\Phi^+_j,\;\; {\mathrm if}\;\; \hat Q_j>0,\nonumber\\
&&\Phi^0_j,\;\; {\mathrm if}\;\; \hat Q_j=0,\nonumber\\
&&\Phi^-_j,\;\; {\mathrm if}\;\; \hat Q_j<0.
\eear
With this classification and knowledge of the sign of the FI term (\ref{fid}),
 we can determine which fields are required for a flat
direction that satisfies (\ref{anomd}).  The statement is as follows:

\noindent Theorem: {\em If $\xi>0$ $(<0)$,
any flat direction must contain at least one of the fields $\Phi^-_j$
$(\Phi^+_j)$}.

Therefore, for each model our strategy is to determine  
$\overline{Q}$,
classify the NAB singlets according to their values of $\hat{Q}_j$
and use the theorem to determine if the model has flat directions formed
out of NAB singlets. If $\xi>0$ ($<0$) and no singlets of the type
$\Phi^-_j$ ($\Phi^+_j$) exist, there is no possibility of forming  $D$-
flat directions out of singlets only which can cancel the FI term. In that
case, fields that transform under non-trivial representations of
NAB groups (either from the hidden sector or the
observable sector) must get VEVs along the flat directions of the model. 

If the $\hat{Q}_j$ classification of the fields is such that $D$- flat
directions can be formed out of the singlet fields, we follow our method 
of \cite{CCEEL2}: first we determine $U(1)_Y$
as a linear combination of the non-anomalous $U(1)$'s, as described in detail
below, select the singlets with zero hypercharge (such that 
${\cal G}_{\rm obs}$ remains unbroken), and check whether fields of the
appropriate type ($\Phi^-_j$ or $\Phi^+_j$) remain with $Y=0$. We then
construct a basis of HIM's, or
equivalently the superbasis of all one dimensional HIM's, that describe the
moduli space of non-anomalous flat directions. The space of flat
directions that also have $D_A=0$ is a subspace of this. It can be formed
by combining the directions in the basis or superbasis, ensuring that
the anomalous charge\footnote{By gauge invariance the $\hat{Q}$ charge of
a HIM coincides with its anomalous charge.} of the resulting direction has
sign opposite to $\xi$.

These $D$- flat directions are then tested for $F$- flatness.  As
discussed in
\cite{CCEEL2}, a subset of the $D$- flat directions can be proved to be
$F$- flat
to all orders in the non-renormalizable superpotential by imposing the
constraints of gauge invariance and knowledge of the superpotential to a given
order.  This subset, referred to as Type-B directions, are those in which one
cannot form total gauge singlet holomorphic operators from the fields that
form the direction\footnote{The presence of the anomalous $U(1)$ is
crucial to this point, as the HIM's associated with good flat directions
are
not invariant under $U(1)_A$.}. Then, for these directions there are only
a finite number of possible terms in the
superpotential which could lift $F$-flatness.  Whether these terms are in
fact
present (i.e., are allowed by string selection rules) is then checked
explicitly.  In contrast, Type-A directions involve fields which can form gauge
group singlets, so that terms which could lift $F$- flatness could occur
to all
orders in the superpotential.  

We will restrict our analysis to type-B flat directions.  Of course,
in doing so we
may leave out some flat directions which are in fact $F$- flat to all
orders, but proving these directions are $F$- flat is a difficult task.
We list the one-dimensional (zero-dimensional after cancelling the
anomalous $D$- term) $D$- flat directions which remain $F$- flat
to all orders, out of which higher dimensional flat directions may be
formed. We also list the number of broken $U(1)$'s for each flat
direction.

\subsection{Hypercharge Determination}

For each model, a viable hypercharge must be determined as a linear
combination of
the nonanomalous\footnote{We do not consider the possibility of $U(1)_Y$
having some component along the generators of hidden sector NAB groups
(broken at some scale). If that breaking is triggered by the FI term, we
should consider flat directions that involve fields in nontrivial
representations of these NAB groups, an analysis which is beyond the scope of
this paper. We also ignore breaking at a lower scale (e.g., radiatively) for
simplicity.  The latter case would also significantly change the picture of SM
gauge coupling unification.  However, the additional matter content in these
models can also modify the gauge unification.  Such phenomenological issues are
also beyond the scope of this paper.}
$U(1)$'s which satisfies the following basic phenomenological criteria:

\begin{itemize}
\item Three generations of quarks and leptons, as well as a pair of
electroweak Higgs doublets with conventional hypercharges.

\item Grouping of all particles with nonzero charge under $SU(3)_C$ or
$U(1)_{EM}$ into mirror pairs, such that mass terms can be generated and
these particles made heavy.
Otherwise, there would necessarily be
exactly massless colored or charged fermions in the theory,
which are clearly 
excluded\footnote{This is the weakest reasonable assumption; one
could make a stronger assumption of no unpaired $SU(2)_L$ chiral states.  Such
additional chiral fermions would be expected to acquire masses when $SU(2)_L$
is broken, and would have important phenomenological consequences,
such as yielding a positive contribution to the electroweak $S$ parameter.
 However, they are not absolutely excluded, so we do not require their
absence as a condition on the hypercharge definition.}.

%\item As the models we presently consider are those with possible singlet
%flat
%directions such that the hidden sector non-Abelian gauge group remains
%unbroken at
%the string 
%scale, we also require that any states which are ``mixed" non-Abelian
%representations of both the observable and hidden sector gauge groups
%come in
%pairs with equal 
%and opposite hypercharges, such that these states can couple to a 
%$Y=0$ non-Abelian singlet which could acquire a string scale VEV. 
%In this way, these mixed fields can decouple from the theory without
%breaking
%the non-Abelian symmetries, and we can
%maintain a distinct barrier between the observable and hidden sectors
%(at least in terms of the non-Abelian gauge groups).  In imposing this
%requirement,
%we are working within a specific phenomenological framework in which we
%wish to
%preserve the distinction between the non-Abelian observable and hidden
%sector
%gauge groups, such that the standard supergravity-mediated mechanism of
%supersymmetry breaking can be implemented\footnote{Alternative
%hypercharge
%definitions in which the chiral states do not pair with each other and
%which
%therefore remain massless until the non-Abelian symmetries are broken
%might be of 
%interest for gauge-mediated supersymmetry breaking.}. 
%Models CHL1, CHL5, and CHL7
%all have such 
%mixed states, and so within our approach this requirement needs to be enforced
%for
%each model.

\end{itemize}

In  general, the number of
fields in each model that are candidates for the observable sector
states is so large that a direct search for hypercharge candidates 
taking all combinations of possible observable sector fields would be
very inefficient.   
Therefore, depending on the number
of nonanomalous $U(1)$'s in the model, we seek the minimal number of constraints
that can determine a hypercharge candidate and then check explicitly if
the above conditions are satisfied.  

In each model considered, there are only  three candidates for the quark
doublet states
$(3,2)$ under $(SU(3)_C,SU(2)_L)$, so
we require
that these fields have the appropriate hypercharges $Y(Q_L)=\frac{1}{6}$.  
However, there are 
generally
more than 6 candidates for the quark singlets $(\bar{3},1)$.  
In some models, there
are $(\bar{3},1)$ states which
are also multiplets under the NAB hidden sector gauge group, so that if
the hidden sector gauge group is broken above the electroweak scale, these
states may also be included in the list of candidates for the quark singlets. 
We scan all combinations of these fields such that we have three
candidates for the right-handed up-type quarks, with hypercharge
$Y(U^c_L)=-\frac{2}{3}$.  If these conditions are not sufficient to
determine $Y$, we also impose similar requirements for the existence of
three right-handed down-type quarks, with $Y(D^c_L)=\frac{1}{3}$.  
We also impose the condition that the
trace of the hypercharge over the remaining $(3,1)$ and $(\bar{3},1)$ 
states (not selected as quarks candidates) is zero. This is a necessary
condition (but not sufficient) for these particles to obtain large masses
and decouple from the low-energy theory.

Once a hypercharge candidate is found that satisfies these conditions
we check explicitly for complete families and pairing of the remaining 
fields in vector-like pairs of equal and opposite electric charge. (Some
hypercharge definitions involve continuous parameters, which can be varied
in the search for pairs).
When several $Y$ definitions exist, one can use the values of $k_Y$ to
discriminate between them \cite{CHL}, choosing the one closer to
$k_Y=5/3$.

\section{Results}

We now apply the methods discussed above to several quasi-realistic
string models taken from the literature, which provide an assorted 
sampling of the different situations one may encounter. Model FNY1 is the one
constructed in ref.~\cite{FNY1}, model AF1 is taken from ref.~\cite{AF1} and
the rest, CHL1 to CHL7, are the models presented in ref.~\cite{CHL}.
The sign of the FI term, proportional to the sign of the trace of
the anomalous charge over all of the fields in the model, is listed in
%Table~\ref{tfiterm}
Table~II
for the models considered.

\subsection{Model FNY1} 
The gauge group of this model is
\be
\{SU(3)_C\times SU(2)_L\}_{\rm obs}\times\{SU(3)\times SU(2)\times
SU(2)\}_{\rm hid}
\times U(1)_A\times U(1)^{11},
\ee
and the particle content includes, besides the MSSM multiplets, additional
chiral superfields:
\bear
&&  1 (\bar{3},1;1,1,1) + 1(3,1;1,1,1) + \nonumber\\
&& 10 (1,2;1,1,1) + 5 (1,1;3,1,1,1) + 5 (1,1;\bar{3},1,1) + \nonumber\\
&& 12 (1,1;1,2,1) + 15 (1,1;1,1,2) + \nonumber\\
&& 57 (1,1;1,1,1)\;\;,
\eear
where the representation under $(SU(3)_C,SU(2)_L;SU(3),SU(2),SU(2))$
is indicated. We list the $U(1)$ charges of the non-Abelian singlets (including
the right-handed leptons) in Table~IIIa. 
The hypercharge definition is given by
\cite{FNY1}
\be
\label{fny1y}
Y=\frac{1}{24}(2Q_1+3Q_2),
\ee
[normalized to give $Y$(quark doublet)$=1/6$], and 
$\overline{Q}$ is given by

\be
\label{fny1qb}
\overline{Q}=\frac{28}{3}(Q_6-Q_{10})-\frac{1}{3}Q_7+28(Q_9-Q_{11}).
\ee
This model has a positive trace of the anomalous charge, so the FI term can be
compensated only along flat directions with negative values of $\hat{Q}$.  
Anticipating the requirement that any such flat direction preserves hypercharge,
we consider only $Y=0$ fields. By inspection, one sees that $S_7$, $S_{20}$,
$S_{36}$, $S_{38}$, and $S_{46}$ have $Y=0$ and $\hat{Q}<0$, so this model
has in principle the possibility of good flat directions.

The basis for the $Y=0$ non-anomalous moduli space is presented in
Table~IV. An element like $M_{31}=\langle 32^2,25^2,6,2,1\rangle$ stands
for the HIM $\Phi_{32}^2\Phi_{25}^2\Phi_6\Phi_2\Phi_1$. There are 41
fields with $Y=0$ and their $41\times 11$ charge
matrix $Q_i^n$ [$i$ is a field index and $n$ a $U(1)_n$ index] has rank 9
[$U(1)_Y$ and $U(1)_{11}$ are zero for this subset of fields]. This
implies that the basis has $32$ elements. In particular, the presence of
primed copies of the fields $S_5$ and $\bar{S}_5$ introduces two 
basis elements ($M_6$ and $M_7$) which are trivially derived from $M_5$. The
last direction in the table, $M_{32}$, has $\hat{Q}<0$ and can be used, by
combining it with other directions, to generate all $\hat{Q}<0$ $D$- flat
directions, i.e., those capable of giving $D_A=0$.

In this model, the superbasis, formed by all one-dimensional flat directions,
has a very large number of elements (ranging in the few thousands), and hence 
the complete determination of the superbasis loses its practical motivation. 
Nevertheless, all the information about the
moduli space of non-anomalous dimensions is already contained in the basis and
any flat direction can be expressed in terms of its elements. This moduli space
contains many directions which are $D_A$ flat (i.e., they have
$\hat{Q}<0$) and also remain $F$- flat. 
In Table~V  we list several examples of such directions that involve
different
numbers of
fields. All of these directions are one-dimensional (before cancelling the
FI term. The free VEV is then fixed to be of order $\xi$) and so break
different numbers of $U(1)$'s, as indicated. As an example, the simplest
direction $R_1$, involving only five different fields, breaks four
non-anomalous
$U(1)$'s. Before compensating the FI term, this direction is
one-dimensional, and
so the VEV's of the five fields are all related (according to the powers
to which they are raised in the associated holomorphic invariant monomial).
After compensating the FI term all the VEV's are then related to
$\xi$ according to
\be
\frac{|\varphi_1|^2}{4}=|\varphi_4|^2=|\varphi_{\bar{6}}|^2=
\frac{|\varphi_7|^2}{2}=\frac{|\varphi_9|^2}{2}
=-\frac{\xi}{\hat{Q}(R_1)}=\frac{\xi}{224}.
\ee
The last directions presented in Table~V are examples that break the maximal
number of $U(1)$'s compatible with unbroken $U(1)_Y$. There is
always another
$U(1)$ besides hypercharge that remains unbroken. 

Some of the listed directions 
do not have any possible superpotential term that could lift the $F$-
flatness (except mass terms
that are absent in superstring models). Many other directions
are lifted already by 
terms in the trilinear superpotential, 
which at that order reads:
\bear
W_3&=&S_4S_{18}S_{38}+S_{4}S_{11}S_{27}+S_{14}S_{46}S_{6}
+S_{8}S_{33}\bar{S}_{6}+S_{1}\bar{S}_{2}\bar{S}_{3}
+\bar{S}_{1}S_{2}S_{3}\nonumber\\
&+&S_{31}S_{35}S_{2}+S_{17}S_{7}\bar{S}_{2}
+S_{26}S_{28}S_{1}+S_{25}S_{32}S_{1}+S_{19}S_{20}\bar{S}_{3}.
\eear

\subsection{Model AF1}
The gauge group is
\be
\{SU(3)_C\times SU(2)_L\}_{\rm obs}\times\{SU(5)\times SU(3)\}_{\rm hid}
\times U(1)_A\times U(1)^9.
\ee
 
Besides the MSSM multiplets, the particle content includes additional
chiral superfields:
\bear
&& 2 (\bar{3},1;1,1) + 2 (3,1;1,1) + 8 (1,2;1,1) + \nonumber\\
&& 4 (1,1;5,1) + 4 (1,1;\bar{5},1) + 8 (1,1;1,3) + \nonumber\\
&& 8 (1,1;1,\bar{3}) + 37 (1,1;1,1)\;\;,
\eear
where the representation under $(SU(3)_C,SU(2)_L;SU(5),SU(3))$ is indicated. In
Table~VIa we list the 40 
NAB singlets (including right-handed leptons) of the
model with their $U(1)$ charges (rescaled by a factor 4 with respect to
ref.~\cite{AF1} to make them integers). 
The hypercharge definition is 
\cite{AF1}
\be
\label{af1y}
Y=\frac{1}{24}(2Q_1+3Q_2),
\ee
[normalized to give $Y$(quark doublet)$=1/6$], and $\overline{Q}$ is
found to be:
\be
\label{af1qb}
\overline{Q}=\frac{1}{2}(-15Q_5+5Q_6+4Q_7-15Q_{8}+Q_{9}).
\ee
In this model, negative values of $\hat{Q}$ are required to
construct flat directions and Table~VIa shows that two $Y=0$ fields have
the correct sign of $\hat{Q}$: $S_7$ and $S_{10}$. In the table,
we also list two other $U(1)$ charges, defined as:
\be
\label{af1qp}
Q'=-26Q_1+3Q_8+5Q_9,\;\;\;\; Q''=15Q_8-Q_9.
\ee
Restricting our attention to $Y=0$ fields, we see that $S_{15}, S_{16}$
and $S_{17}$ are the only fields with non-zero $Q'$ charges, and all 
of them are negative and equal. This implies that no HIM built of
non-Abelian $Y=0$ singlets can contain these three fields, so they do not
appear in this type of flat direction and we can ignore them in the
following. We are then left with (note that all the fields have a
mirror copy) $N^*=13\times 2$ ($Y=0$) fields
which have zero charge under three independent $U(1)$'s: $Y, Q'$ and
$Q''$.  The rank of the
non-anomalous charge matrix for this subset of
fields is then equal to 6. Consequently, we expect a basis of
non-anomalous flat directions composed of 20 elements. Two such bases are
presented in Table~VII. The first is constructed in such a way as to
minimize the number (and power) of the fields entering the basis elements.
The second contains a sub-basis  of $\hat{Q}<0$ elements. All
type-B $\hat{Q}<0$ directions can be obtained by combining the
elements of this sub-basis.

The superbasis, containing all one-dimensional non-anomalous flat
directions of the model, 
can be readily constructed and is presented in tables VIIIa and VIIIb. 
It contains a total of
$157\times 2$ elements (every flat direction is accompanied by another
formed by the mirror copies of the fields in the first one). The
superscript 0 labels $\hat{Q}=0$ directions and the rest have
$\hat{Q}=-60$ ($+60$ for the mirror direction not written).
In total, 123 have the correct sign of $\hat{Q}$ to compensate the
FI term, and thus are true flat directions.
These elements ($P_\alpha$ with $\alpha=1,...,123$) are the building
blocks of all type-B $D$- flat directions.

Many of the $D$- flat directions built out of the $P_\alpha$'s
will be lifted by $F$- terms. The superpotential of the model involving
NAB singlets only is \cite{AF1}, up to the fourth order:
\bear
W_2&=&0,\vspace{0.1cm}\label{af1w2}\\
W_3&=&
S_3\overline{S}_4 S_{12}+\overline{S}_3S_4 \overline{S}_{12}\nonumber\\
&+&\overline{S}_{11}(S_5\overline{S}_8 +S_6\overline{S}_9
+S_7\overline{S}_{10} +S_{12}S_{13})
\nonumber\\
&+&S_{11}(\overline{S}_5S_8 +\overline{S}_6S_9
+\overline{S}_7S_{10} +\overline{S}_{12}\overline{S}_{13})\label{af1w3}\\
&+&\overline{S}_{1}S_{2}S_{3},\nonumber\vspace{0.1cm}\\
W_4&=&0\label{af1w4}.
\eear
The different Yukawa couplings, of order $g$, are not indicated
explicitly. The mirror copy of the term $\overline{S}_{1}S_{2}S_{3}$
is absent, forbidden by world-sheet selection rules, as are all
terms quadratic and quartic in the fields, that would otherwise be allowed
by gauge symmetries.

Knowledge of $W$ up to fourth order terms is nearly all that is needed to 
determine which $P_\alpha$'s are also $F$- flat. It turns out that only
10 of them remain flat to all orders in the presence of $F$- terms.
The particular direction 
\be
\label{nuis}
P=R_{81}=\langle 1^4,\bar{2},\bar{4},5,6,7,\overline{13}^2\rangle,
\ee
however, requires knowledge of up to sixth order terms in $W$, as it can
be
lifted if the terms
\bear
W_B^{(5)}&=&S_1\overline{S}_2\overline{S}_4S_{11}\overline{S}_{13},\nonumber\\
W_B^{(6)}&=&S_1\overline{S}_2\overline{S}_4
(S_5\overline{S}_8+S_6\overline{S}_9+S_7\overline{S}_{10})\overline{S}_{13},
\eear
do appear in the actual superpotential.

While gauge invariant terms like $W_B^{(5)}$ and $W_B^{(6)}$ 
are expected to appear in field-theoretic models unless additional
symmetries are enforced {\it ad hoc}, 
the situation is different in string models.
Beyond spacetime symmetries, stringy superpotential terms 
must always satisfy a set of world-sheet selection rules \cite{wsrules}, 
as we discussed in section two.
In this case, neither $W_B^{(5)}$ nor any of the three $W_B^{(6)}$ terms 
meet world-sheet selection requirements and, therefore, are
eliminated from the superpotential. 
While the above terms are gauge group invariants, their corresponding
five- and six-point string amplitudes are zero as a result of
string effects, as explained below.

That $W_B^{(5)}$ does not survive in the superpotential
is relatively easy to demonstrate:
fields $S_1$, $\overline{S}_2$, and $\overline{S}_4$ are Ramond fields,
that is, they originate in twisted world-sheet supersymmetric sectors of the 
string model. 
In contrast $S_{11}$ and $\overline{S}_{13}$ come from the untwisted 
\NS  world-sheet supersymmetric sector of the model. 
A picture changed \cite{FMS} set of $3+K$ (with $K\ge 1$) states can 
form an invariant under the global $U(1)_{N=2}$ symmetry 
of the $N=2$ world-sheet supersymmetry only if no more
than $K-1$ fields are of the \NS type \cite{RT}.
Hence, a stringy candidate $W_5$ term with
two \NS fields has a vanishing five-point amplitude, since such a term
is not invariant under $U(1)_{N=2}$.

The three $W_B^{(6)}$ terms cannot be discarded so easily, for they 
each contain exactly one \NS field while up to two are allowed. 
For these terms we 
must also examine their additional global $U(1)$ world-sheet
charges and/or Ising field correlation functions.  
From this we can demonstrate that 
none of the three terms appear in the superpotential: 
first, we can show that the pairs of fields that
distinguish the three $W_B^{(6)}$ terms, 
i.e., $S_5\overline{S}_8$, $S_6\overline{S}_9$, and
$S_7\overline{S}_{10}$, 
 all originate in the
same twisted subsector of the model. The three 
pairs all follow the same
basic pattern with regard to their global world-sheet charges and 
Ising fields. 

Next, we choose the fields 
$S_1$, $\overline{S}_2$, $\overline{S}_4$ to be the three
that are not picture changed in any of the three terms. 
%(which we are free to do
%since any $(3+K)$-point string amplitude is invariant under 
%different choices of which set of $K$ fields are chosen to be picture-changed. 
Then 
$\overline{S}_2$, $\overline{S}_4$ contribute to the 
six-point string amplitude two distinct  
Ising twist field correlators $\langle\sigma_i^{+}\sigma_i^{-}\rangle$ (i=1,2),
associated, respectively, with two non-chiral
Ising world-sheet fermions, that we will denote $(f_1(z),\ov{f}_1(\ov{z}))$ and
$(f_2(z),\ov{f}_2(\ov{z}))$ .    
Both before and after all possible picture changing options
(including those that lead to a $U(1)_{N=2}$ conserving term) 
are performed on the respective sets of fields, 
$\{S_5\overline{S}_8\overline{S}_{13}\}$, 
$\{S_6\overline{S}_9\overline{S}_{13}\}$, and
$\{\overline{S}_{13} S_7\overline{S}_{10}\}$, 
we find that there are no further contributions to  
either non-chiral Ising fermion's correlation functions.
Since $\langle \sigma_i^{+}\sigma_i^{-} \rangle= 0$, the entire
six-point string amplitudes are zero and the terms are removed from
the superpotential. Thus, while we should generically expect flatness
of the particular direction (\ref{nuis}) to be broken by the
appearance of $W_B^{(5)}$ and $W_B^{(6)}$ in a generic field-theoretic
model, direction (\ref{nuis}) remains flat in this string model 
due to additional world-sheet selection rules. 
%This is a typical example
%of the effect that stringy origins, versus field-theoretic origins, 
%can have on a model.

Table~IX lists all one-dimensional (zero-dimensional after cancelling the
FI term) $P_{\alpha}$ directions that remain $F$- flat to all orders.
Other multi-dimensional flat directions can be built by multiplying
these together in different combinations. Some exceptions arise if one
combines $P_{\alpha}$'s in such a way that the fields $S_{12}$ and $S_{13}$
(or $S_{13}$ and $\overline{S}_{13}$)
take VEV's simultaneously. In the first case, $F$- terms are
generated at the Yukawa level [see eq.~(\ref{af1w3})] and lift the direction
while in the second case the flat direction would be of type A (it contains
the
HIM $S_{13}\ov{S}_{13}$) and knowledge of the superpotential to all orders
would be required to ensure that no lifting term $(S_{13}\ov{S}_{13})^n$
appear.

\subsection{Model CHL1}

This model has the gauge group
\be
\{SU(3)_C\times SU(2)_L\}_{\rm obs}\times\{SU(2)_2\times
SU(2)^2\}_{\rm hid}
\times U(1)_A\times U(1)^{13},
\ee
and the particle content includes,\footnote{The particle contents
of models CHL1 though CHL3 are presented and discussed in another
paper in this series, \cite{GC}.}  
besides the MSSM multiplets, 
the additional chiral superfields:
\bear
&& 5 (3,1;1,1,1) + 5 (\bar{3},1;1,1,1) + \nonumber\\
&& 4 (1,1;2,1,2) + 4 (1,1;2,2,1) + 4 (1,1;1,2,2) + \nonumber\\
&& 2 (1,2;2,1,1) + 2 (1,1;3,1,1) + \nonumber\\
&& 16(1,1;2,1,1) + 10(1,1;1,2,1) + 10(1,1;1,1,2) + \nonumber\\
&& 14(1,2;1,1,1) + 84(1,1;1,1,1) \, ,
\eear
where the representation under 
$(SU(3)_C,SU(2)_L;SU(2)_2,SU(2)^2)$
is indicated. There are two bi-doublets $(2,2)$ under 
$SU(2)_L\times SU(2)_2$ that
mix the observable and hidden sectors, so that only if these fields
get string scale masses 
%[or $SU(2)_2$ is broken] 
are the two sectors really
separated. In Table~X we list the 87  non-Abelian singlets 
 of the model with their $U(1)$ charges (including right-handed leptons). 

 In the CHL models \cite{CHL}, the trace of the anomalous charge is
negative, and thus we require positive values of $\hat{Q}$.
$\overline{Q}$ is given by
\be 
\label{chl1qb}
\overline{Q}=\frac{32}{33}Q_6-\frac{64}{473}Q_7 
+\frac{48}{43}Q_8-\frac{64}{129}Q_9\\ 
+ \frac{176}{6063}Q_{10}-\frac{288}{2491}Q_{11} 
+\frac{40}{2067}Q_{12}+\frac{11}{39}Q_{13}.
\ee
As shown in Table~X, there are four NAB-singlet fields with a positive
value of $\hat{Q}$ so that, in principle,
good flat directions may be formed.  However, a viable definition of
hypercharge also must be determined, to see whether flat directions
exist that preserve $U(1)_Y$.

 In Model CHL1, the search  yielded no acceptable hypercharge
which had three families,
had exotic $SU(3)_C$ triplet pairing, and the possibility of the
decoupling of 
mixed observable and hidden
bi-doublet states. However, we can impose a weaker vector pairing
requirement allowing for the possible breaking of part of the hidden
sector NAB gauge groups (i.e., allowing the pairing 
between fields belonging to different representations of those groups, with
typical masses at the breaking scale\footnote{An alternative
possibility is that the shift to a SUSY preserving vacuum requires
non-zero VEV's of fields that transform under a non-trivial representation
of some NAB hidden sector group. In either case preservation of $U(1)_Y$
requires the breaking of the NAB group. We focus on the first possibility,
as we are interested in exploring vacuum shifts involving NAB-singlet
fields only.  In this case, the NAB group can be broken at a lower scale,
unrelated to the anomalous $U(1)$ breaking.}).  In that case, 
several definitions of $Y$ could be found. One particular example 
(which gives the lowest value of $k_Y$, $k_Y=8/3$) is
\bear
\label{chl1y}
Y&=&\frac{1}{8}(Q_1-Q_3)\nonumber\\
&&+\frac{1}{12}\left[
-\frac{1}{11}Q_6+\frac{45}{946}Q_7 
+\frac{15}{172}Q_8+ \frac{5}{43}Q_9
+ \frac{45}{8084}Q_{10}+\frac{69}{4982}Q_{11} 
+\frac{1}{106}Q_{12}\right].
\eear
The hypercharges of NAB-singlet fields according to this definition are
also listed in Table~X: a total of 33 fields are left with $Y=0$ (and
sufficient candidates for singlet leptons with $Y=1$ appear). In addition,
the charges of the fields under the linear combination 
\be
\label{chl1qp}
Q'=Q_1+Q_2,
\ee
are also given.
After examination of the $Q'$ charges of the $Y=0$ fields, we conclude that
the field $S_{33}$ cannot enter any flat direction (that preserves $Y$
and is built of NAB singlets only). The only field with $\hat{Q}>0$ is
$S_{14}$, while $S_{29}$, $S_{34}$, $S_{38}$, $S_{47}$, $S_{49}$,
$S_{55}$ and $S_{63}$ have negative $\hat{Q}$. 

In this model, the number of elements of the superbasis is large, making
its complete determination unwieldy.   
Therefore, we use a basis that describes the space of $Y=0$
non-anomalous $D$- flat
directions, which is presented in Table~XI.   
It contains 21 elements, corresponding to the fact that there are 32 ($Y=0$)
NAB singlets left after removing $S_{33}$, and the $U(1)$ 
charge matrix for these 32 fields has rank 11 [13 minus 2 $U(1)$'s always
unbroken: $Y$ and $Q'$]. 
All the basis elements have $\hat{Q}$ either zero
or negative, while a positive value would be required to cancel the FI
term. However, this does not necessarily imply that there are no flat
directions with $\hat{Q}>0$ (in contrast to the case with the elements of
the superbasis, in which an element with the right sign of $\hat{Q}$ is
required for $D_A$- flat directions). 
By the definition of the basis, any $D$- flat direction $P$ can be written
in the form
\be
\label{pn}
P^n=\Pi_i M_i^{n_i}=\Pi_i\left[\Pi_j\varphi_j^{m_{ij}}\right]^{n_i},
\ee
where $n$, $n_i$ and $m_{ij}$ are integer numbers, with $n,m_{ij}>0$, and 
the $n_i$ not necessarily positive. 
The only condition for $P^n$ to be acceptable is
that all the fields appearing in it are raised to a positive power, which
is not equivalent to requiring positive $n_i$'s. From eq.~(\ref{pn}) it
follows
\be
\hat{Q}(P^n)=\sum_i n_i \hat{Q}(M_i)=\sum_i n_i\left[
\sum_j m_{ij} \hat{Q}(\varphi_j)\right],
\ee
opening the possibility of obtaining $\hat{Q}(P^n)>0$ via $n_i<0$ for
some $\hat{Q}(M_i)<0$. Whether this can
be realized depends on the details
of the model.  In the following we illustrate how the
knowledge of the basis can be used to prove general statements about
flat directions.  

In this model and with the definition of $Y$ given above, the only field with
positive $\hat{Q}$ is $S_{14}$, so, if a $P$ exists for which $\hat{Q}(P)>0$,
its definition (\ref{pn}) in
terms of the basis elements must include $M_{19}$ and/or $M_{21}$ raised
to some positive power, because these are the only basis elements that
contain $S_{14}$. In both elements, $S_{14}$ appears in combination with
$S_{49}$, which neutralizes its positive $\hat{Q}$. We are then
forced to include in (\ref{pn}) some element which contains also $S_{49}$
but appears raised to a negative power so as to cancel the power of
$S_{49}$ in the final expression for $P$. The only basis element available
for this purpose is $M_{20}$, but it cannot have a
negative power in (\ref{pn}) because it contains the field $S_{34}$ which
appears only in this basis element and then, the final expression for $P$
would contain a negative power of $S_{34}$, which cannot be accepted.
This proves that no $D$- flat direction exists with $\hat{Q}>0$.

If we do not fix the hypercharge definition from the beginning and include
all the NAB singlets in the analysis, the moduli space of non-anomalous
flat directions is larger and is described by the basis presented in
Table~XII. The number of basis elements is 74 [87 (fields)- 13(rank)] 
and only one of them has $\hat{Q}$ non-zero (and negative). In this case,
however, flat directions with $\hat{Q}>0$ exist. As an example,
\be
P=\langle
1^7,\ov{5}^7,6^{15},7^6,14^6,16^4,20^{17},21^{22},27^2,42^2,75^5
\rangle,
\ee
has $\hat{Q}=64\times 12$. Its expression in terms of the basis elements
is
\be
P=\frac{1}{M_{73}^3}\times\left\{
\frac{M_{42}^{17}M_{21}^{3}M_{74}^{6}M_{22}^{3}M_{31}^{5}
M_{1}^{3}M_{39}^{5}M_{13}^{2}M_{57}^{2}M_{34}M_{7}^{5}
}{M_{16}^{5}M_{62}^{2}M_{8}M_{6}^{8}M_{14}M_{9}M_{56}^{5}M_{5}M_{12}^{17}}
\right\}.
\ee
This gives an explicit example of a model in which the basis has no
element with good $\hat{Q}$ but good $\hat{Q}$ $D$- flat directions
exist\footnote{
This is in contrast with the situation for the fields themselves: no good
flat directions can exist if the model does not contain fields with the
appropriate value of $\hat{Q}$.}.
 
\subsection{Model CHL2} 
The gauge group of this model is
\be
\{SU(3)_C\times SU(2)_L\}_{\rm obs}\times\{SO(7)\times
SU(2)_2^2\}_{\rm hid}
\times SU(2)^4 
\times U(1)_A\times U(1)^7,
\ee
and the particle content \cite{GC} includes additional chiral
superfields:
\bear
&& 3 (1,1;8,1,2,1,1,1,1) + (1,1;8,1,1,1,2,1,1) + (1,1;7,3,1,1,1,1,1) +
\nonumber\\
&&(1,1;1,2,3,1,1,1,1) + (1,1;7,2,1,1,1,1,1) + (1,1;7,1,1,1,1,1,1) +
\nonumber\\
&& 2 (1,1;1,2,2,2,1,1,1) + 2 (1,1;1,2,2,1,1,1,2) + (1,1;1,2,1,1,1,2,2) + 
\nonumber\\
&& (1,1;1,2,1,2,2,1,1) + 3 (1,1;1,1,1,1,2,2,1) + 2 (1,1;1,1,1,1,2,1,2) + 
\nonumber\\
&& 3 (1,1;1,1,1,2,1,1,2) + (1,1;1,1,1,1,1,2,2) + (1,1;1,1,1,1,2,1,2) +  
\nonumber\\
&& 3 (1,1;1,1,1,2,1,2,1) + 3 (1,1;1,3,1,1,1,1,1) + (1,1;1,1,3,1,1,1,1) + 
\nonumber\\
&& 10 (1,1;1,2,1,1,1,1,1) + 4 (1,2;1,2,1,1,1,1,1) + (1,2;1,2,2,1,2,1,1) +
\nonumber\\
&& (1,2;1,1,1,1,1,2,2) + 4 (1,2;1,1,1,1,1,1,1) +  3 (\bar{3},1;1,2,1,1,1,1,1)+ 
\nonumber\\
&& (3,1;1,2,1,1,1,1,1) + (3,1;1,1,1,2,2,1,1) + 2 (\bar{3},1;1,1,1,1,1,1,1) + 
\nonumber\\
&& 16 (1,1;1,1,1,1,1,1,1)\;\;,
\eear
where the representation under 
$(SU(3)_C,SU(2)_L;SO(7),SU(2)^2_2,SU(2)^4)$
is indicated. In Table~XIII we list the 19 
non-Abelian singlets (including right-handed leptons) of the model with their $U(1)$ charges. 

$\overline{Q}$ is given by
\be
\label{chl2qb}
\overline{Q}=\frac{1}{3}(28Q_4+Q_5).
\ee
Inspection of the list of singlets in this model shows that all of the
fields either have zero or negative values of $\hat{Q}$, so that in this
model non-Abelian fields are required for a flat direction. We conclude
that the shifting to a SUSY preserving vacuum is necessarily accompanied
by the spontaneous breaking of some non-Abelian group.
To be more precise, we find that only a single hidden sector non-Abelian
field has a positive $\hat{Q}$ value. This field is a doublet under a
level-one $SU(2)$ and both level-two $SU(2)$'s, which implies that in the
least these three gauge groups must be broken if the SM is to survive after
anomaly cancellation.

For completeness, we construct the basis of non-anomalous flat
directions built out of NAB singlets, which can be useful for more general
discussions when non-singlet fields are also included. In Table~XIII
we also list the charges under the linear combination of $U(1)$'s
defined as:
\be
\label{chl2qp}
Q'=6Q_1+4Q_4-Q_5+Q_7.
\ee
We see that $S_{11}$, $S_{12}$, $S_{13}$
and $S_{14}$ are the only fields with non-zero $Q'$ charges, and all 
of them are positive and equal. This implies that no HIM built of
non-Abelian singlets can contain these four fields, so they do not
appear in this type of flat direction and we can ignore them in the
following discussion. We are then left with 15 fields
which have zero charge under (\ref{chl2qp}).
The rank of the
non-anomalous charge matrix for this subset of
fields is then equal to 6 and we expect a basis of
non-anomalous flat directions composed of 9 elements. Such a basis is
presented in Table~XIV.

\subsection{Model CHL3} 
The gauge group is
\be
\{SU(3)_C\times SU(2)_L\}_{\rm obs}\times\{SO(5)^2\times SU(2)_2\}_{\rm hid}
\times U(1)_A\times U(1)^{11},
\ee
and the particle content \cite{GC} includes the additional chiral superfields:
\bear
&& 4 (\ov{3},1;,1,1,1) + 4 (3,1;1,1,1) + 8 (1,2;1,1,1) + \nonumber\\
&& 4 (1,1;1,1,3) + 16 (1,1;1,4,1) + 8 (1,1;4,1,1) + \nonumber\\
&& 2 (1,1;1,5,1) + 2 (1,1;5,1,1) + (1,1;5,5,1) + \nonumber\\
&& 4 (1,1;4,1,2) + 2 (1,1;1,4,2) + 2 (1,2;4,1,1) + \nonumber\\
&& 76 (1,1;1,1,1)\;\;,
\eear
where the representation under $(SU(3)_C,SU(2)_L;
SO(5),SO(5),SU(2)_2)$ is indicated. In Table~XV
we list the 79 
non-Abelian singlets (including right-handed leptons) of the
model with their $U(1)$ charges. 
 
In this model, $\overline{Q}$ is :
\be
\label{chl3qb}
\overline{Q}=-\frac{1}{5}(6Q_8+9Q_{10}-4Q_{11}).
\ee
As in CHL2, all of the fields have either zero or negative values of
$\hat{Q}$ so that  non-Abelian fields are required to cancel the FI term
along any flat direction.
We present in Table~XVI the basis of non-anomalous
$D$- flat directions for non-Abelian singlets.  Imposing 11 non-anomalous
$D$- term conditions on 79 fields leads to a moduli space of 68 dimensions.

\subsection{Model CHL4} 
The gauge group of this model is
\be
\{SU(3)_C\times SU(2)_L\}_{\rm obs}\times\{SU(4)_2\times SU(2)_2\}_{\rm hid}
\times U(1)_A\times U(1)^6,
\ee
and the particle content beyond the MSSM consists of the chiral superfields:
\bear
&&12 (1,2;1,1) + 2(3,1;1,1) + 2(\bar{3},1;1,1) + \nonumber\\
&&2 (1,1;6,2) + 2 (1,1;\bar{4},2) + 4 (1,1;6,1) + \nonumber\\
&&14 (1,1;4,1) + 10 (1,1;\bar{4},1) + 3 (1,1;1,3) + 8 (1,1;1,2)+\nonumber\\
&& 48 (1,1;1,1)\;\;,
\eear
where the representation under $(SU(3)_C,SU(2)_L;
SU(4)_2,SU(2)_2)$ is indicated. In Table~XVII
we list the 51 
non-Abelian singlets (including right-handed leptons) of the
model with their $U(1)$ charges. 
 
Phenomenological considerations lead to the hypercharge definition
\cite{CHL}
\be
\label{chl4y}
Y=\frac{1}{24}\left(-\frac{3}{5}Q_1+\frac{27}{80}Q_2+\frac{3}{10}Q_3
+\frac{1}{6}Q_4-\frac{1}{6}Q_5+\frac{5}{48}Q_6\right),
\ee
[normalized to give $Y$(quark doublet)$=1/6$]. As previously explained,
$k_Y$ for this definition of hypercharge can be readily calculated
using the universal GS relation with only the knowledge of the charges of
the massless spectrum of the string model. In this way, we find $k_Y=35/12$,
a factor of 2 larger than the $k_Y$ quoted in \cite{CHL}, and thus greater
than $5/3$. This discrepancy by a factor of 2 affects all other
determinations of $k_Y$ presented in \cite{CHL} and seems to eliminate the
examples with $k_Y<5/3$.
 
In this model, $\overline{Q}$ is particularly simple:

\be
\label{chl4qb}
\overline{Q}=-\frac{1}{2}Q_6.
\ee
As is shown in Table~XVII, $\hat{Q}$ is negative or zero for all of the 
non-Abelian singlet fields, while ${\mathrm Tr}Q_A<0$, so it is
not possible to form a good flat direction without utilizing the
non-Abelian fields. From Table~XVII one concludes also that alternative
definitions of $Y$ do not change this situation.

Even if non-Abelian fields would have to take non-zero VEV's along any true
flat direction, it may be necessary to give VEV's to singlets as well.
For this reason we present, as in previous cases, the basis of
non-anomalous $D$- flat directions for non-Abelian $Y=0$ singlets in
Table~XVIII.
There are 17 $Y=0$ fields plus 3 copies, and 5 non-anomalous $U(1)$'s besides
hypercharge so that the number of elements of the basis is 12 (plus 3 more
involving copies of fields, that are not shown).

\subsection{Model CHL5}
This model, with gauge group
\be
\{SU(3)_C\times SU(2)_L\}_{\rm obs}\times\{SU(4)_2\times SU(2)_2\}_{\rm hid}
\times U(1)_A\times U(1)^{6},
\ee
was already considered in ref.~\cite{CCEEL2} to which we refer the reader for
further details. 
The model contains a NAB singlet with $\hat{Q}>0$ that can appear in several
flat directions with good $\hat{Q}>0$.
A total of 5 one-dimensional type-B flat directions were found that could
be used as building blocks for directions which are $D$- and $F$- flat to all
orders. 
The phenomenological analysis of the model along some of these directions
will be presented elsewhere \cite{CCEELW}.

\subsection{Model CHL6} 
The gauge group of this model is
\be
\{SU(3)_C\times SU(2)_L\}_{\rm obs}\times\{SU(2)_2^2\}_{\rm hid}
\times U(1)_A\times U(1)^{10},
\ee
and it includes the additional chiral superfields:
\bear
&& 3 (\bar{3},1;1,1) + 3 (3,1;1,1) +6 (1,2;1,1) + \nonumber\\
&& 4 (1,2;2,1) + 26 (1,1;2,1) + 40 (1,1;1,2) + \nonumber\\
&& 1 (1,1;3,3) + 1 (1,1;3,1) + 1 (1,1;1,3) + 49 (1,1;1,1)\;\;,
\eear
where the representation under $(SU(3)_C,SU(2)_L;SU(2)_2,SU(2)_2)$ is
indicated. In Table~XIX we list the 52 
non-Abelian singlets (including right-handed leptons) of the model 
with their $U(1)$ charges. 

$\overline{Q}$ is given by
\be
\label{chl6qb}
\overline{Q}=-\frac{1}{5}(6Q_7+9Q_9-4Q_{10}).
\ee
Inspection of the list of singlets in this model shows that all of the
fields have zero values of $\hat{Q}$, except for $S_{24}$ with
$\hat{Q}=-144$, which is of sign opposite to the FI term, so that
hidden (or observable) non-Abelian fields must take non-zero VEV's along
a flat direction.

The basis of non-anomalous flat directions is presented in Table~XX.
The number of elements is obtained subtracting from the total number of fields
($N$=52) the rank of the charge matrix (10) giving a dimension equal to 42 
(in this case, we explicitly include primed fields).

\subsection{Model CHL7} 
The gauge group of this model is
\be
\{SU(3)_C\times SU(2)_L\}_{\rm obs}\times\{SO(7)_2\times SU(2)_2\}_{\rm hid}
\times U(1)_A\times U(1)^8,
\ee
and the particle content includes, besides the MSSM multiplets, additional
chiral superfields:
\bear
&&14 (1,2;1,1) + 6 (3,1;1,1) + 4 (\bar{3},1;1,1) + \nonumber\\
&&1 (1,2;1,2) + 1 (\bar{3},1;1,2) +4 (1,1;8,1) + \nonumber\\
&&1 (1,1;7,1)  + 4 (1,1;1,3) + 31 (1,1;1,2)+\nonumber\\
&& 94 (1,1;1,1)\;\;,
\eear
where the representation under $(SU(3)_C,SU(2)_L;
SU(4)_2,SU(2)_2)$ is indicated. In Table~XXI
we list the 81 non-Abelian singlets (including right-handed leptons. The
13 fields with zero charges under all of the Abelian groups are not
listed) of the model with their $U(1)$ charges. 

$\overline{Q}$ is given by
\be
\label{chl7qb}
\overline{Q}=-4Q_3-4Q_4+Q_6-2Q_7-\frac{4}{3}Q_{8},
\ee
and there are fields with positive $\hat{Q}$ that can
in principle form a good flat direction.  Once again, a viable hypercharge
must be determined to ensure that such good flat directions preserve the
SM gauge group.

Upon further inspection of the list of states, one can see
that in this model there is an additional mixed state which is a color
antitriplet and 
hidden sector doublet $(\bar{3},1;1,2)$. Thus, to enforce
triplet pairing (i.e., to avoid  an exactly massless colored fermion),the
hidden sector $SU(2)$ must be broken to have the appropriate number of
degrees of freedom.  

Even allowing for the breaking of any group [except $SU(3)_C\times
U(1)_{EM}$], no definition of $Y$ as a linear combination of the
non-anomalous $U(1)$'s exists that gives full vector pairing of
the additional multiplets present in this model.  As explained in
section~III.B we did not consider the
possibility that the hypercharge definition  involves the $U(1)$'s that
arise from the breaking of the hidden sector NAB gauge group.
In case such a possibility can indeed be realized it can be useful to know
the basis of non-anomalous flat directions of all NAB singlets, which is
presented in Table~XXII.

\section{Conclusions}

We have applied the strategy developed in \cite{CCEEL2} for the
classification of flat directions to several quasi-realistic models
[with an anomalous $U(1)$] taken from the literature
\cite{FNY1,AF1,CHL}. The results are summarized in
Table~II and offer a survey of the different possibilities that can be
encountered in this type of analyses.

\begin{itemize}
\item In some cases (CHL2, CHL3, CHL4 and CHL6) it is possible to show 
that the FI
term $\xi$ cannot be compensated by giving VEV's to non-Abelian singlet fields
only. The technical reason is that no such fields exist with $\hat{Q}$ of
sign opposite to that of $\xi$ (or of ${\mathrm Tr}Q_A$, as listed in
Table~II). This
holds irrespective of the definition of hypercharge and is the reason why we
did not search for a viable $Y$ in some of these models (marked '?' in the
corresponding column of Table~II).
For these models we thus conclude that the vacuum shifting triggered by the
FI term is necessarily accompanied by the reduction of the rank of the
non-Abelian group. The analysis of flat directions involving fields in non
trivial representations of the non-Abelian groups is beyond the scope of this
paper. Such an analysis could address the issue of whether the SM gauge group
will be necessarily broken. 

\item In other models (e.g., CHL1), even if they contain non-Abelian
singlets with
good $\hat{Q}$ (and possibly flat directions) 
it can happen that, after
determining a viable definition of hypercharge $Y$,
 no flat directions
remain that preserve $U(1)_Y$. 
This can happen because no $Y=0$ fields are left
with good $\hat{Q}$ or, in a more subtle way, because, even if they exist
the charge structure of the fields conspire to produce directions
with only the wrong sign of $\hat{Q}$ (this was the case of CHL1 with the
particular choice of $Y$). 
If that $U(1)_Y$ is to survive unbroken in the low-energy
effective theory, the vacuum restabilization must be accompanied
by the breaking of some non-Abelian gauge group.

\item Models exist (CHL7) for which no definition of hypercharge is
phenomenologically viable (the presence of massless charged particles in the 
spectrum cannot be avoided). 

\item Other models (FNY1, AF1, CHL5) are more successful and have both a viable
hypercharge and hypercharge-preserving flat directions. In these
cases, our conservative aim is to
classify all such directions that can be proven to be flat to all orders.
This can be done by constructing the superbasis that contains all
one-dimensional directions which are $D$- flat for the non-anomalous $U(1)$'s.
Of these directions, only those that carry an anomalous charge of sign
opposite to $\xi$ are also $D_A$-flat. The $F$- flatness to all orders of
these particular directions (and some combinations of them) can be assessed
by knowing the superpotential up to a finite order, and thus those directions
which are both $D$- and $F$- flat can in principle be classified.

This program can be readily completed \cite{CCEEL2} for model CHL5, which
contains a small number of all-order flat directions. 
The number of flat directions increases for model AF1 
and hence the superbasis is large, but
the number of flat directions which are $F$- flat to all orders is still
relatively small.
However, for model FNY1 the superbasis is too large to be of practical use.
This model simply contains too many flat directions to give a complete
classification (even though in principle this can be done). 

For these models we find and list several flat directions that break
different numbers of $U(1)$'s. In general, not only $U(1)_Y$ survives
down to low-energies but other $U(1)$'s remain unbroken
at the string scale. The fate of these additional Abelian factors
depends on the details of the model and was addressed, for example, in
refs.~\cite{CDEEL,CCEEL1}.
\end{itemize}

Having found a classification of the different vacua to which a particular
model can relax,
the next step is to analyze the spectrum, gauge group and 
superpotential of the resulting model, and  investigate their phenomenological
consequences. This analysis for model CHL5
is currently under investigation \cite{CCEELW}.

\acknowledgments
This work was supported in part by U.S. Department of Energy Grant No. 
DOE-EY-76-02-3071. J.R.E. thanks the Physics Department of the
University of Pennsylvania for hospitality and financial support during
the later stages of this work. We thank Jing Wang for helpful discussions.
\newpage

%*********************************************************************
\def\NPB#1#2#3{{\it Nucl.\ Phys.}\/ {\bf B#1} (#2) #3}
\def\PLB#1#2#3{{\it Phys.\ Lett.}\/ {\bf B#1} (#2) #3}
\def\PRD#1#2#3{{\it Phys.\ Rev.}\/ {\bf D#1} (#2) #3}
\def\PRL#1#2#3{{\it Phys.\ Rev.\ Lett.}\/ {\bf #1} (#2) #3}
\def\PRT#1#2#3{{\it Phys.\ Rep.}\/ {\bf#1} (#2) #3}
\def\PTP#1#2#3{{\it Prog.\ Theor.\ Phys.}\/ {\bf#1} (#2) #3}
\def\MODA#1#2#3{{\it Mod.\ Phys.\ Lett.}\/ {\bf A#1} (#2) #3}
\def\IJMP#1#2#3{{\it Int.\ J.\ Mod.\ Phys.}\/ {\bf A#1} (#2) #3}
\def\nuvc#1#2#3{{\it Nuovo Cimento}\/ {\bf #1A} (#2) #3}
\def\RPP#1#2#3{{\it Rept.\ Prog.\ Phys.}\/ {\bf #1} (#2) #3}
\def\etal{{\it et al\/}}
%*********************************************************************

\newpage

\def\pnas{$\phantom{\rm NAS}$}
\def\phna{$\phantom{\rm HSA}$}
\def\pona{$\phantom{\rm OSA}$}
\def\p#1{$\phi_{#1}$}
\def\bp#1{$\bar{\phi}_{#1}$}
\def\pp#1{$\phi^{'}_{#1}$}
\def\bpp#1{$\bar{\phi}^{'}_{#1}$}
\def\hf#1{$H_{#1}$}
\def\bh#1{$\bar{H}_{#1}$}
\def\v#1{$V_{#1}$}
\def\bv#1{$\bar{V}_{#1}$}

\begin{center}
\begin{tabular}{|c|c|ccccc|ccccc|ccccc|}
\hline\hline
Field&$\fer^{12}$&$S_{12}$&$y^1$&$\w^1$&$y^2$&$\w^2$
                 &$S_{34}$&$y^3$&$\w^3$&$y^4$&$\w^4$ 
                 &$S_{56}$&$y^5$&$\w^5$&$y^6$&$\w^6$\\
\hline\hline
$H_{39,\, ferm\, (\mh)}$&$\half$   
                          &$0  $&$0  $&$\mh$  &$0$  &$\sm$
                          &$0  $&$0  $&       &$\sm$&$0$  
                          &$\mh$&$0  $&$0$    &     &      \\  
\hline
$H_{37,\, ferm\, (\mh)}$&$\half$   
                          &$0  $&$0  $&$\mh$  &$0$  &$\sm$
                          &$0  $&$0  $&       &$\sm$&$0$  
                          &$\mh$&$0  $&$0$    &     &      \\  
\hline
$H_{32,\, bos\, (-1)}$ &$0$  
                          &$\ph$&$\mh$&$0$    &$\sp$&$0$  
                          &$0$  &$0$  &       &$0$  &$0$
                          &$\ph$&$\sp$&$0$    &     &      \\
\hline
$H_{30,\, bos\, (-1)}$ &$0$  
                          &$\ph$&$\mh$&$0$    &$\sp$&$0$  
                          &$0$  &$0$  &       &$0$  &$0$
                          &$\ph$&$\sp$&$0$    &     &      \\
\hline
net charges            &$0$
                          &$+1$ &$-1$&$-1$    &$0$  &$0$  
                          &$0$  &$0$  &       &$0$  &$0$
                          &$0$  &$0$  &$0$    &     &      \\
\hline
$T^{-1}_{3/2}$ term    &$0$
                          &$-1$ &$+1$&$+1$    &$0$  &$0$  
                          &$0$  &$0$  &       &$0$  &$0$
                          &$0$  &$0$  &$0$    &     &      \\
\hline\hline
\end{tabular}
\end{center}
\noindent Table~I: 
Picture-changing example in a $W_4$ term
$H_{39} H_{37} H_{32} H_{30}$ in \cite{FNY1}.   
These fields are identified in our Tables~IIIa and IIIb as
$S_{43}$, $S_{28}$, $S_{24}$, and $S_{7}$.
The ``net charges'' row contains the charge vector formed from
the four canonical $H$ fields that must be cancelled by picture-changing
affects.

\begin{center}
\begin{tabular}{|c||c|c|c|c|c|}
\hline\hline
Model & ${\rm Tr}\, Q_A$ & $Y$ def. & $D$-flat & $D$-flat ($Y=0$)& $F$-flat 
\\ 
\hline\hline
FNY1      & 1344   & Y &  Y & Y   & Y   \\
AF1       &  720   & Y &  Y & Y   & Y   \\
CHL1      &$-$3072 & Y &  Y & N   & $-$ \\
CHL2      &$-$2688 & ? &  N & $-$ & $-$ \\
CHL3      &$-$3456 & ? &  N & $-$ & $-$ \\
CHL4      &$-$2016 & Y &  N & $-$ & $-$ \\
CHL5      &$-$1536 & Y &  Y & Y   & Y   \\
CHL6      &$-$3456 & Y &  N & $-$ & $-$ \\
CHL7      &$-$768  & N &  Y & $-$ & $-$ \\
\hline\hline
\end{tabular}
\end{center}
\noindent Table~II: 
Summary of results for the different models considered. 
The second column gives the total trace of the anomalous $U(1)$. The third
shows whether a viable definition of hypercharge exists (a question mark
indicates that no answer to that question is required to proceed with the
analysis). The fourth and fifth columns relate to the existence of
$D$-flat directions, in the latter case imposing also that $Y=0$ along them.
The last column reports the existence of $D$- flat directions which
remain $F$- flat.

\begin{center}
\begin{tabular}{|c|rrrrrrrrrrrr|r|r|}
\hline\hline
NA        &&&&&&&&&&&&&&\\ 
Singlet     &$Q_A$&$Q_1$&$Q_2$&$Q_3$&$Q_4$&$Q_5$&$Q_6$
                  &$Q_7$&$Q_8$&$Q_9$&$Q_{10}$&$Q_{11}$
                  &$Q_Y$&$\hat{Q}$\\
\hline\hline
$S_1$$\surd$&$-$36&   0&   0&   4&  16&   0&   0& 108&   0&   0&   0&   0&0&0\\
$S_2$$\surd$&$-$28&   0&   0&   4& $-$32&   0&   0&  84&   0&   0&   0&
0&0&0\\
$S_{3}$ $\surd$& $-$8&   0&   0&   0&  48&   0&   0&  24&   0&   0&   0&
0&0&0\\
$S_{4}$ $\surd$& $-$8&   0&   0&   0&   0&   0&  $-$8&$-$200&   0&   0&   0&
0&0&0\\
$S_{5}{}^({}'{}^)$ $\surd {}^({}'{}^)$
                      &  0&   0&   0&   0&   0&   0&   0&   0& 4& 0& 0& 0&0&0\\
$S_{6}$ $\surd$&  0&   0&   0&   0&   0&   8&   0&   0&   0&   0&   0&  0&0&0\\
$S_{7}$&$-$24&  $-$3&   2&  $-$2& $-$17&   2&   0& $-$96&   0&   1&  $-$3&
0&0&$-$112\\
$S_8$ &$-$16&   3&  $-$2&   0&  $-$3&   4&   2& $-$64&   2&  $-$1&   3&
0&0&0\\
$S_{9}$&$-$12&   3&  $-$2&  $-$6& $-$15&   2&   4& $-$20&   0&  $-$1&   3&
0&0&0\\
$S_{10}$ &$-$12&   0&   4&  $-$2&  10&   2&  $-$2& $-$20&   2&   2&   0&
2&1&0\\
$S_{11}$ &$-$12&   3&  $-$2&   0&  $-$3&   0&   6&  36&  $-$2&  $-$1&   3&
0&0&0\\
$S_{12}$ & $-$8&   3&   2&   2&  11&   2&   4& $-$32&   2&  $-$1&  $-$3&
$-$2&1&$-$112\\
$S_{13}$ & $-$8&   3&   2&   2& $-$13&   2&   0&$-$144&  $-$2&  $-$1&  $-$3&
2&1&0\\
$S_{14}$ & $-$8&   3&  $-$2&   0&  $-$3&  $-$4&  $-$2& 136&  $-$2&  $-$1&   3&
0&0&112\\
$S_{15}$ & $-$8&   0&  $-$4&  $-$2&  10&  $-$2&   2&  80&   2&   2&   0&
2&$-$1&0\\
$S_{16}$ & $-$8&   0&  $-$4&  $-$2& $-$14&  $-$2&  $-$2& $-$32&   2&   2&   0&
2&$-$1&0\\
$S_{17}$& $-$4&   3&  $-$2&   6& $-$15&  $-$2&   0& 180&   0&  $-$1&   3&
0&0&112\\
$S_{18}$& $-$4&   3&  $-$2&   0&  $-$3&   0&   2& 236&   2&  $-$1&   3&
0&0&112\\
$S_{19}$& $-$4&   0&   0&   0&  24&   0&   0&  12&   2&  $-$2&   6&   0&0&112\\
$S_{20}$& $-$4&   0&   0&   0&  24&   0&   0&  12&  $-$2&   2&  $-$6&
0&0&$-$112\\
$S_{21}$ & $-$4&   0&   4&  $-$2& $-$14&   2&   2&  68&   2&   2&   0&  2&1&0\\
$S_{22}$ & $-$4&  $-$3&  $-$2&  $-$2&  13&   2&   0&$-$156&  $-$2&   1&   3&
2&$-$1&0\\
$S_{23}$ & $-$4&   3&   2&   2&  11&  $-$2&  $-$4&  68&   2&  $-$1&  $-$3&
2&1&112\\
$S_{24}$&  0&  $-$3&   2&   2&  11&  $-$2&   0& 168&   0&  $-$3&  $-$3&
0&0&112\\
$S_{25}$& 20&   0&   0&  $-$2&  $-$8&  $-$2&   2&  $-$4&   0&   2&   6&
0&0&0\\
$S_{26}$& 20&  $-$3&   2&  $-$6&  $-$9&   2&   0& 108&   0&   1&  $-$3&
0&0&0\\
$S_{27}$& 20&  $-$3&   2&   0&   3&   0&   2& 164&   2&   1&  $-$3&   0&0&0\\
$S_{28}$& 16&   3&  $-$2&   2&  $-$7&  $-$2&   0&$-$216&   0&  $-$1&   3&
0&0&0\\
$S_{29}$& 16&   0&   0&   4&   4&   0&   0& $-$48&   2&   2&   6&   0&0&0\\
$S_{30}$& 16&   0&   0&   4&   4&   0&   0& $-$48&  $-$2&  $-$2&  $-$6&
0&0&0\\
$S_{31}$& 16&   0&   0&  $-$2&  16&   2&   2&   8&   0&   2&   6&   0&0&0\\
$S_{32}$& 16&   0&   0&  $-$2&  $-$8&   2&  $-$2&$-$104&   0&  $-$2&  $-$6&
0&0&0\\
$S_{33}$ & 16&  $-$3&   2&   0&   3&   4&  $-$2&  64&  $-$2&   1&  $-$3&
0&0&0\\
$S_{34}$& 12&   6&   4&   2& $-$10&   2&   2&  20&   0&   0&   0&   0&2&0\\
$S_{35}$& 12&   0&   0&  $-$2&  16&  $-$2&  $-$2& $-$92&   0&  $-$2&  $-$6&
0&0&0\\
$S_{36}$& 12&  $-$3&   2&   6&  $-$9&  $-$2&   4& $-$92&   0&   1&  $-$3&
0&0&$-$112\\
$S_{37}$ & 12&   0&  $-$4&   2& $-$10&  $-$2&   2&  20&   2&  $-$2&
0&$-$2&$-$1&0\\
$S_{38}$ & 12&  $-$3&   2&   0&   3&   0&   6& $-$36&  $-$2&   1&
$-$3&0&0&$-$112\\
$S_{39}$&  8&   6&   4&  $-$4&   2&   0&   0& $-$24&  $-$2&   0&   0&   0&2&0\\
$S_{40}$&  8&   6&  $-$4&  $-$4&   2&   0&   0& $-$24&  $-$2&   0&   0&
0&0&0\\
\hline
\end{tabular}
\end{center}

\begin{center}
\begin{tabular}{|c|rrrrrrrrrrrr|r|r|}
\hline
NA        &&&&&&&&&&&&&&\\ 
Singlet     &$Q_A$&$Q_1$&$Q_2$&$Q_3$&$Q_4$&$Q_5$&$Q_6$
                  &$Q_7$&$Q_8$&$Q_9$&$Q_{10}$&$Q_{11}$
                  &$Q_Y$&$\hat{Q}$\\
\hline\hline
$S_{41}$&  8&   6&  $-$4&   2& $-$10&  $-$2&  $-$2& $-$80&   0&   0&
0&0&0&0\\
$S_{42}$&  8&   6&   4&   2&  14&  $-$2&   2&  32&   0&   0&   0&   0&2&0\\
$S_{43}$&  8&   3&  $-$2&  $-$2&  13&   2&   0& 144&   0&   3&   3&   0&0&0\\
$S_{44}$ &  8&  $-$3&  $-$2&  $-$2& $-$11&  $-$2&  $-$4&  32&   2&   1&   3&
2&$-$1&112\\
$S_{45}$ &  8&  $-$3&  $-$2&  $-$2&  13&  $-$2&   0& 144&  $-$2&   1&   3&
$-$2&$-$1&0\\
$S_{46}$ &  8&  $-$3&   2&   0&   3&  $-$4&   2&$-$136&   2&   1&  $-$3&
0&0&$-$112\\
$S_{47}$ &  8&   0&   4&   2& $-$10&   2&  $-$2& $-$80&   2&  $-$2&   0&
$-$2&1&0\\
$S_{48}$ &  8&   0&   4&   2&  14&   2&   2&  32&   2&  $-$2&   0& $-$2&1&0\\
$S_{49}$&  4&   6&  $-$4&   2&  14&   2&  $-$2& $-$68&   0&   0&   0&   0&0&0\\
$S_{50}$ &  4&   0&  $-$4&   2&  14&  $-$2&  $-$2& $-$68&   2&  $-$2&   0&
$-$2&$-$1&0\\
$S_{51}$ &  4&   3&   2&   2& $-$13&  $-$2&   0& 156&  $-$2&  $-$1&  $-$3&
$-$2&1&0\\
$S_{52}$ &  4&  $-$3&  $-$2&  $-$2& $-$11&   2&   4& $-$68&   2&   1&   3&
$-$2&$-$1&$-$112\\
\hline\hline
\end{tabular}
\end{center}
\noindent Table IIIa: List of non-Abelian singlet fields and their $U(1)$ gauge 
charges for model FNY1, with hypercharge and $\hat{Q}=Q_A-\overline{Q}$ as defined in
eqs.~(\ref{fny1y}) and (\ref{fny1qb}) respectively. A $\surd$ indicates
the presence of another field with equal and opposite $U(1)$ charges,
while a $'$ indicates the presence of another field with identical $U(1)$
charges.

\def\ibf{$\bar{f}$}
\def\msp{$\sigma^{+}$}
\def\msm{$\sigma^{-}$}
\def\BS{\bar{S}}
\def\xre{${\re_1}:{\re_2}= $} 
\begin{center}
\begin{tabular}{|c|ccccccc|cccccc|}
\hline\hline

                &\xre&   3:6 &  4:20 &  5:11 &  9:12 & 10:19 & 15:18 &  7:38 &  8:44 & 13:40 & 14:46 & 16:41 & 17:47\\   
\hline
Singlet $S_i$   &    &     Q &       &       &       &       &       &     I &       &       &       &       &      \\ 
\hline\hline
$ S_{ 1}$       &    &   .5  &       &       &   .5  &       &  $-$.5  &       &       &       &       &       &      \\  
$\BS_{1}$       &    &   .5  &       &       &   .5  &       &  $-$.5  &       &       &       &       &       &      \\  
$ S_{ 2}$       &    &   .5  &       &       &  $-$.5  &       &   .5  &       &       &       &       &       &      \\  
$\BS_{2}$       &    &   .5  &       &       &  $-$.5  &       &   .5  &       &       &       &       &       &      \\  
$ S_{ 3}$       &    &  $-$.5  &       &       &   .5  &       &   .5  &       &       &       &       &       &      \\  
$\BS_{ 3}$      &    &  $-$.5  &       &       &   .5  &       &   .5  &       &       &       &       &       &      \\  
$ S_{ 4}$       &    &  $-$.5  &       &       &   .5  &       &   .5  &       &       &       &       &       &      \\  
$\BS_{ 4}$      &    &  $-$.5  &       &       &   .5  &       &   .5  &       &       &       &       &       &      \\  
$ S_{ 5}$       &    &   .5  &       &       &  $-$.5  &       &   .5  &       & \ibf  &       &       &       &      \\  
$\BS_{5}$       &    &   .5  &       &       &  $-$.5  &       &   .5  &       & \ibf  &       &       &       &      \\  
$\S'_{5}$       &    &   .5  &       &       &   .5  &       &  $-$.5  & \ibf  &       &       &       &       &      \\  
$\BS'_{5}$      &    &   .5  &       &       &   .5  &       &  $-$.5  & \ibf  &       &       &       &       &      \\  
$ S_{6}$        &    &  $-$.5  &       &       &   .5  &       &   .5  &       &       &       &       &       &      \\  
$\BS_{6}$       &    &  $-$.5  &       &       &   .5  &       &   .5  &       &       &       &       &       &      \\  
$ S_{ 7}$       &    &       &   .5  &       &   .5  &       &       & \msm  &       &       &       & \msp  &      \\  
$ S_{ 8}$       &    &   .5  &       &       &       &  $-$.5  &       &       &       &       & \msm  & \msp  &      \\  
$ S_{ 9}$       &    &       &       &  $-$.5  &       &       &   .5  &       & \msm  & \msm  &       &       &      \\  
$ S_{10}$       &    &       &   .5  &       &   .5  &       &       &       & \msp  &       &       &       & \msm \\  
$ S_{11}$       &    &   .5  &       &       &       &  $-$.5  &       &       &       &       & \msp  & \msm  &      \\  
$ S_{12}$       &    &       &   .5  &       &   .5  &       &       &       & \msm  &       &       & \msm  &      \\  
$ S_{13}$       &    &       &       &  $-$.5  &       &       &   .5  & \msp  &       & \msp  &       &       &      \\  
$ S_{14}$       &    &   .5  &       &       &       &  $-$.5  &       &       &       & \msm  &       &       & \msm \\  
$ S_{15}$       &    &       &   .5  &       &   .5  &       &       &       & \msm  &       &       &       & \msm \\  
$ S_{16}$       &    &       &       &   .5  &       &       &   .5  & \msm  &       &       & \msm  &       &      \\ 
$ S_{17}$       &    &       &  $-$.5  &       &   .5  &       &       & \msm  &       &       &       & \msp  &      \\  
$ S_{18}$       &    &   .5  &       &       &       &  $-$.5  &       &       &       & \msp  &       &       & \msp \\  
$ S_{19}$       &    &   .5  &       &       &       &   .5  &       &       &       &       & \msp  &       & \msm \\  
$ S_{20}$       &    &   .5  &       &       &       &  $-$.5  &       &       &       &       & \msp  &       & \msm \\  
$ S_{21}$       &    &       &       &   .5  &       &       &   .5  & \msp  &       &       & \msm  &       &      \\  
$ S_{22}$       &    &       &       &  $-$.5  &       &       &   .5  & \msm  &       & \msp  &       &       &      \\  
$ S_{23}$       &    &       &  $-$.5  &       &   .5  &       &       &       & \msm  &       &       & \msm  &      \\  
$ S_{24}$       &    &       &   .5  &       &   .5  &       &       & \msm  &       &       &       & \msp  &      \\  
$ S_{25}$       &    &       &       &   .5  &       &       &   .5  &       & \msm  &       & \msp  &       &      \\  
$ S_{26}$       &    &       &       &   .5  &       &       &   .5  &       & \msp  & \msm  &       &       &      \\  
$ S_{27}$       &    &   .5  &       &       &       &   .5  &       &       &       &       & \msp  & \msm  &      \\  
$ S_{28}$       &    &       &       &  $-$.5  &       &       &   .5  &       & \msp  & \msm  &       &       &      \\  
$ S_{29}$       &    &   .5  &       &       &       &   .5  &       &       &       & \msp  &       & \msp  &      \\  
$ S_{30}$       &    &   .5  &       &       &       &  $-$.5  &       &       &       & \msm  &       & \msm  &      \\  
\hline
\end{tabular}
\end{center}

\begin{center}
\begin{tabular}{|c|ccccccc|cccccc|}
\hline
                &\xre&   3:6 &  4:20 &  5:11 &  9:12 & 10:19 & 15:18 &  7:38 &  8:44 & 13:40 & 14:46 & 16:41 & 17:47\\   
\hline
Singlet  $S_i$  &    &     Q &       &       &       &       &       &     I &       &       &       &       &      \\ 
\hline\hline
$ S_{31}$       &    &       &   .5  &       &   .5  &       &       & \msp  &       &       &       &       & \msp \\  
$ S_{32}$       &    &       &       &  $-$.5  &       &       &   .5  &       & \msm  &       & \msp  &       &      \\  
$ S_{33}$       &    &   .5  &       &       &       &   .5  &       &       &       &       & \msm  & \msp  &      \\  
$ S_{34}$       &    &       &   .5  &       &   .5  &       &       & \msm  &       &       &       &       & \msp \\  
$ S_{35}$       &    &       &  $-$.5  &       &   .5  &       &       & \msp  &       &       &       &       & \msp \\  
$ S_{36}$       &    &       &   .5  &       &   .5  &       &       & \msp  &       &       &       & \msp  &      \\  
$ S_{37}$       &    &       &  $-$.5  &       &   .5  &       &       &       & \msm  &       &       &       & \msm \\  
$ S_{38}$       &    &   .5  &       &       &       &   .5  &       &       &       & \msp  &       &       & \msp \\  
$ S_{39}$       &    &   .5  &       &       &       &   .5  &       &       &       & \msm  &       & \msm  &      \\  
$ S_{40}$       &    &   .5  &       &       &       &   .5  &       &       &       & \msp  &       & \msp  &      \\  
$ S_{41}$       &    &       &   .5  &       &   .5  &       &       & \msp  &       &       &       &       & \msp \\  
$ S_{42}$       &    &       &       &   .5  &       &       &   .5  &       & \msp  &       & \msp  &       &      \\  
$ S_{43}$       &    &       &       &  $-$.5  &       &       &   .5  &       & \msp  & \msm  &       &       &      \\  
$ S_{44}$       &    &       &  $-$.5  &       &   .5  &       &       &       & \msp  &       &       & \msm  &      \\  
$ S_{45}$       &    &       &       &   .5  &       &       &   .5  & \msm  &       & \msp  &       &       &      \\  
$ S_{46}$       &    &   .5  &       &       &       &   .5  &       &       &       & \msm  &       &       & \msm \\  
$ S_{47}$       &    &       &  $-$.5  &       &   .5  &       &       &       & \msp  &       &       &       & \msm \\  
$ S_{48}$       &    &       &       &  $-$.5  &       &       &   .5  & \msp  &       &       & \msm  &       &      \\  
$ S_{49}$       &    &       &       &   .5  &       &       &   .5  &       & \msm  &       & \msp  &       &      \\  
$ S_{50}$       &    &       &       &  $-$.5  &       &       &   .5  & \msm  &       &       & \msm  &       &      \\  
$ S_{51}$       &    &       &       &   .5  &       &       &   .5  & \msp  &       & \msp  &       &       &      \\  
$ S_{52}$       &    &       &   .5  &       &   .5  &       &       &       & \msp  &       &       & \msm  &      \\  
\hline\hline
\end{tabular}
\end{center}
\noindent Table~IIIb: 
The non-gauge worldsheet charges of the 60 NAB singlets in model FNY1. 
${\re_1}$ and ${\re_2}$ specify the two real fermions $\psi_{\re_1}$ and $\psi_{\re_2}$ 
comprising either a complex left-moving fermion when $\re_2 \leq 20$, 
or a non-chiral Ising fermion when $\re_2 > 20$. 
A global $U(1)$ charge $Q$ carried 
by a singlet $S_i$ is listed in the column of the
complex worldsheet fermion associated with the charge. 
Likewise, a conformal field $I\in \{ f, \bar{f}, \sp, \sm \}$ of a
non-chiral Ising fermion 
carried by a singlet is listed in the column of
the appropriate Ising fermion. 

\vskip 1.truecm

\begin{center}
\begin{tabular}{|l|r||l|r|}
\hline\hline
BASIS      &   $\hat{Q}$   &   & $\hat{Q}$ \\
\hline\hline
$M_{1}=\langle 1,\ov{1}\rangle$ &0 &
	$M_{17}=\langle 33,8,\ov{6}\rangle$ &0 \\
$M_{2}=\langle 2,\ov{2}\rangle$ &0 &
	$M_{18}=\langle 20,19,\ov{3}\rangle$ &0 \\
$M_{3}=\langle 3,\ov{3}\rangle$ &0 & 
	$M_{19}=\langle 49,25,24,7\rangle$ &0 \\
$M_{4}=\langle 4,\ov{4}\rangle$ &0 & 
	$M_{20}=\langle 43,28,24,7\rangle$ &0 \\
$M_{5}=\langle 5,\ov{5}\rangle$ &0 & 
	$M_{21}=\langle 40,29,24,7\rangle$ &0 \\
$M_{6}=\langle 5', \ov{5}\rangle$ &0 & 
	$M_{22}=\langle 41,31,24,7\rangle$ &0 \\
$M_{7}=\langle 5, \ov{5}'\rangle$ &0 & 
	$M_{23}=\langle 46,40,33,19,2\rangle$ &0 \\
$M_{8}=\langle 6,  \ov{6}\rangle$ &0 & 
	$M_{24}=\langle 38,27,14,8,4\rangle$ &0 \\	
$M_{9}=\langle 32,25,1\rangle$ &0 & 
	$M_{25}=\langle 43,36,24,9,4\rangle$ &0 \\
$M_{10}=\langle 28,26,1\rangle$ &0 &
	$M_{26}=\langle  31,30,14,7,5\rangle$ &0 \\
$M_{11}=\langle \ov{3},\ov{2},1\rangle$ &0 &
	$M_{27}=\langle 35,29,18,7,\ov{5}\rangle$&0 \\
$M_{12}=\langle 35,31,2\rangle$ &0 &
	$M_{28}=\langle 30,27,25,9,4,1\rangle$ &0 \\
$M_{13}=\langle 27,11,4\rangle$ &0 &
	$M_{29}=\langle 33,32,29,9,\ov{6},1 \rangle$ &0 \\
$M_{14}=\langle 38,18,4\rangle$ &0 &
	$M_{30}=\langle  49,46,26,19,\ov{5},2\rangle$ &0 \\
$M_{15}=\langle 46,14,6\rangle$ &0 &
	$M_{31}=\langle  35^2,25^2,6,2,1\rangle$ &0 \\
$M_{16}=\langle 17,7,\ov{2} \rangle$ &0 &
	$M_{32}=\langle 38,14,9,7,5,4,1^2\rangle$ &$-$112 \\
\hline
\hline
\end{tabular}
\end{center}
\noindent Table IV: Basis of the moduli space of
$Y=0$ non-anomalous $D$-flat directions of model FNY1.

\begin{center}
\begin{tabular}{|l|c|c|c|}
\hline\hline
FLAT DIRECTION    &   Dim.  & $\#$ $U(1)$'s& $-\hat{Q}/112$  \\
\hline\hline
$R_{1}=\langle 1^4,4,\ov{6},7^2,9^2\rangle$    & 0 & 4 & 2\\
\hline
$R_{2}=\langle 1^3,2,\ov{6},7^4,19^2,40^2\rangle$    & 0 & 5 & 2\\
$R_{3}=\langle 1^2,7^2,9,14,29,35\rangle$    & 0 & 5 & 1\\
\hline
$R_{4}=\langle 1^2,\ov{4},\ov{6},7^6,14^2,19^2,49^2\rangle$    & 0 & 6 &2\\
$R_{5}=\langle 1,\ov{5},\ov{6},7^3,18,19,49\rangle$    & 0 & 6 &1\\
$R_{6}=\langle 1,5,7^3,14^2,19,38,49\rangle$    & 0 & 6 &1\\
$R_{7}=\langle 1^3,\ov{6},7^4,9,14,19,49\rangle$    & 0 & 6 &2\\
\hline
\hline
\end{tabular}
\end{center}
\noindent Table V: List of some type-B $D$-flat directions that are
$F$-flat to all orders for the model FNY1. The dimension of the
direction, after
cancellation of the Fayet-Iliopoulos term, is indicated in the second
column. The third column gives the number of non-anomalous $U(1)$'s
broken along the flat direction and the fourth lists the corresponding
values of $-\hat{Q}/112$.

\begin{center}
\begin{tabular}{|c|rrrrrrrrrr|r|r|r|r|}
\hline\hline
NA          &&&&&&&&&&&&&&\\ 
Singlet    &$Q_A$&$Q_1$&$Q_2$&$Q_3$&$Q_4$&$Q_5$&$Q_6$
                  &$Q_7$&$Q_8$&$Q_9$
                  &$Q_Y$&$\hat{Q}$&$Q'$&$Q''$\\
\hline\hline
$S_1$$\surd$&  $-$16&  0&  0&  0&  4&  0&  0& $-$8&  0&  0&0&0&0&0\\
$S_2$$\surd$&  $-$10& $-$3&  2&  2& $-$2&  0&  0& $-$5& $-$1&$-$15&0&0&0&0\\
$S_3$$\surd$&   $-$6&  3& $-$2& $-$2&  6&  0&  0& $-$3&  1& 15&0&0&0&0\\
$S_4$$\surd$&   $-$6&  3& $-$2&  2& $-$6&  0&  0& $-$3&  1& 15&0&0&0&0\\
$S_5$$\surd$&   $-$4&  0&  0&  4&  0&  0& $-$8&  8&  0&  0&0&0&0&0\\
$S_6$$\surd$&   $-$4&  0&  0&  4&  0&  4&  4&  8&  0&  0&0&0&0&0\\
\hline
\hline
\end{tabular}
\end{center}
\noindent Table VIa: List of non-Abelian singlets for model AF1, with
hypercharge
and $\hat{Q}=Q_A-\overline{Q}$ as defined in eqs.~(\ref{af1y}) and
(\ref{af1qb}), respectively. Also shown are the $Q'$ and $Q''$ charges
defined in eq.~(\ref{af1qp}).

\begin{center}
\begin{tabular}{|c|ccccccc|cccccc|}
\hline\hline
                &\xre&   3:6 &  4:17 &  8:14 &  9:12 & 10:19 & 15:18 &  5:43 &  7:38 & 11:45 & 13:40 & 16:41 & 20:48\\   
\hline
Singlet $S_i$   &    &     Q &       &       &       &       &       &     I &       &       &       &       &      \\ 
\hline\hline
$ S_{1 }$       &    &       &       &       &       &       &   .5  & \msm  & \msm  & \msm  & \msm  &       &      \\ 
$\BS_{1 }$      &    &       &       &       &       &       &   .5  & \msm  & \msp  & \msm  & \msp  &       &      \\ 
$ S_{2 }$       &    &       &       &       &   .5  &       &       & \msm  & \msp  &       &       & \msm  & \msp \\ 
$\BS_{2 }$      &    &       &       &       &   .5  &       &       & \msp  & \msp  &       &       & \msp  & \msp \\ 
$ S_{3 }$       &    &   .5  &       &       &       &       &       &       &       & \msm  & \msp  & \msm  & \msp \\ 
$\BS_{3 }$      &    &   .5  &       &       &       &       &       &       &       & \msp  & \msp  & \msm  & \msm \\
\hline
\hline
\end{tabular}
\end{center}
\noindent Table~VIb: 
The non-gauge worldsheet charges of the 40 NAB singlets in AF1.

\vskip .3truecm
\begin{center}
\begin{tabular}{|l|r||l|r|}
\hline\hline
BASIS A      &   $\hat{Q}$   &BASIS B   & $\hat{Q}$ \\
\hline\hline
 $M_1=\langle 1,\bar{1}\rangle$ &0 & 
	$P_1=\langle 1^3,5,6,10,\ov{13}\rangle$ &$-$60 \\
$M_{2}=\langle 2,\bar{2}\rangle$&0&
	$P_2=\langle 1^3,5,7,9,\ov{13} \rangle$&$-$60\\
$M_{3}=\langle 3,\bar{3}\rangle$ &0&
	$P_3=\langle 1^3,5,9,10,12 \rangle$ &$-$60\\
$M_{4}=\langle 4,\bar{4}\rangle$&0&
	$P_4=\langle 1^3,6,7,8,\ov{13}\rangle$&$-$60\\
$M_{5}=\langle 5,\bar{5}\rangle$ &0& 
	$P_5=\langle 1^2,2,4,5,9,10\rangle$ & $-$60\\
$M_{6}=\langle 6,\bar{6}\rangle$    &0  &
	$P_6=\langle 1^3,\bar{3},4,5,9,10\rangle$ &$-$60 \\
$M_{7}=\langle 7,\bar{7}\rangle$ &0&
	$P_7=\langle 1^3,\bar{5},6,7,8^2,12\rangle$ &$-$60 \\
\hline
\hline
\end{tabular}
\end{center}
\noindent Table VII: Two different bases of the moduli space of
non-anomalous $D$-flat directions of the model AF1. 

\begin{center}
\begin{tabular}{|l|l|l|}
\hline\hline
$R_{1}^0=\langle 1,\ov{1}\rangle$  &
	$R_{41}^0=\langle \ov{1},2,4,\ov{6},9,13\rangle$  &
		$R_{81}=\langle 1^4,\ov{2},\ov{4},5,6,7,\ov{13}^2\rangle$\\
$R_{2}^0=\langle 2,\ov{2}\rangle$  &
	$R_{42}=\langle 1^2,2,4,7,8,9\rangle$  &
		$R_{82}=\langle 1^2,2,4,5,6,10,\ov{11}\rangle$ \\	
$R_{3}^0=\langle 3,\ov{3}\rangle$  & 
	$R_{43}^0=\langle \ov{1},2,4,\ov{7},10,13\rangle$  &
		$R_{83}=\langle 1^2,2,4,5,7,9,\ov{11}\rangle$ \\
$R_{4}^0=\langle 4,\ov{4}\rangle$  & 
	$R_{44}=\langle 1^3,\ov{3},4,5,9,10\rangle$ &
		$R_{84}=\langle 1^2,2,4,6,7,8,\ov{11}\rangle$  \\
$R_{5}^0=\langle 5,\ov{5}\rangle$  & 
	$R_{45}=\langle 1^3,\ov{3},4,6,8,10\rangle$  &
		$R_{85}=\langle 1^2,2,4,8,9,10,11\rangle$ \\
$R_{6}^0=\langle 6,\ov{6}\rangle$  & 
	$R_{46}=\langle  1^3,\ov{3},4,7,8,9\rangle$  &
		$R_{86}=\langle 1,2^2,4^2,8,9,10,13\rangle$  \\
$R_{7}^0=\langle 7,\ov{7}\rangle$  & 
	$R_{47}=\langle  1^3,\ov{5},6,7,8^2,12\rangle$  &
		$R_{87}=\langle 1^3,\ov{3},4,\ov{5},6,7,8^2\rangle$ \\
$R_{8}^0=\langle 8,\ov{8}\rangle$  & 
	$R_{48}=\langle 1^3,5,\ov{6},7,9^2,12\rangle$ &
		$R_{88}=\langle 1^3,\ov{3},4,5,\ov{6},7,9^2\rangle$  \\
\hline\hline
\end{tabular}
\end{center}
\noindent Table VIIIa: Superbasis of the moduli space of
non-anomalous $D$-flat directions of model AF1.

\begin{center}
\begin{tabular}{|l|l|l|}
\hline\hline
$R_{121}=\langle 2^3,3^3,7,8,9,11,\ov{13}\rangle$  &
	$R_{134}=\langle  2^3,4^3,\ov{5}^2,6,8^3,10,13^2\rangle$  &
		$R_{147}=\langle  2^3,4^3,6,\ov{7}^2,8,10^3,13^2\rangle$\\
$R_{122}=\langle 2^3,3^3,8,9,10,11,12\rangle$  &
	$R_{135}=\langle  2^3,4^3,5,\ov{6}^2,9^3,10,13^2\rangle$  &
		$R_{148}=\langle  2^3,4^3,6,7,8,\ov{11},\ov{12}^2\rangle$ \\	
$R_{123}=\langle 2^3,3^3,8,9,10,11^2,\ov{13}\rangle$  & 
	$R_{136}=\langle  2^3,4^3,5,6,10,\ov{11},\ov{12}^2\rangle$  &
		$R_{149}=\langle  2^3,4^3,6,7,8,\ov{11}^3,13^2\rangle$ \\
$R_{124}=\langle 2^3,3^3,8,9,10,12^2,13\rangle$  & 
	$R_{137}=\langle  2^3,4^3,5,6,10,\ov{11}^3,13^2\rangle$ &
		$R_{150}=\langle  2^3,4^3,6,7,8,\ov{12}^3,\ov{13}\rangle$  \\
$R_{125}=\langle 2^3,4^3,\ov{5},6,7,8^2,\ov{12}^2\rangle$  & 
	$R_{138}=\langle  2^3,4^3,5,6,10,\ov{12}^3,\ov{13}\rangle$  &
		$R_{151}=\langle  2^3,4^3,\ov{6},8,9^2,10,13^2\rangle$ \\
$R_{126}=\langle 2^3,4^3,\ov{5}^3,6,7,8^4,13^2\rangle$  & 
	$R_{139}=\langle   2^3,4^3,\ov{5}^2,7,8^3,9,13^2\rangle$  &
		$R_{152}=\langle  2^3,4^3,6,8,10,\ov{11}^2,13^2\rangle$  \\
$R_{127}=\langle  2^3,4^3,5,\ov{6},7,9^2,\ov{12}^2\rangle$  & 
	$R_{140}=\langle   2^3,4^3,5,\ov{7}^2,9,10^3,13^2\rangle$  &
		$R_{153}=\langle  2^3,4^3,\ov{7},8,9,10^2,13^2\rangle$ \\
$R_{128}=\langle  2^3,4^3,5,\ov{6}^3,7,9^4,13^2\rangle$  & 
	$R_{141}=\langle  2^3,4^3,5,7,9,\ov{11},\ov{12}^2\rangle$ &
		$R_{154}=\langle  2^3,4^3,7,8,9,\ov{11}^2,13^2\rangle$  \\
\hline\hline
\end{tabular}
\end{center}
\noindent Table VIIIb: Superbasis of the moduli space of
non-anomalous $D$-flat directions of model AF1 (continued).

\vskip 1.truecm

\begin{center}
\begin{tabular}{|l|c|c|}
\hline\hline
FLAT DIRECTION    &   Dim.  & $\#$ $U(1)$'s  \\
\hline\hline
$R_{1}=\langle 1^3,5,6,10,\ov{13}\rangle$    & 0 & 4 \\
$R_{2}=\langle 1^3,5,7, 9,\ov{13}\rangle$    & 0 & 4 \\
$R_{3}=\langle 1^3,5,9,10,    12 \rangle$    & 0 & 4 \\
$R_{4}=\langle 1^3,6,7, 8,\ov{13}\rangle$    & 0 & 4 \\
$R_{5}=\langle 1^3,6,8,10,    12 \rangle$    & 0 & 4 \\
$R_{6}=\langle 1^3,7,8, 9,    12 \rangle$    & 0 & 4 \\
$R_{7}=\langle 1^2,2,4,5,9,10    \rangle$    & 0 & 5 \\
$R_{8}=\langle 1^2,2,4,6,8,10    \rangle$    & 0 & 5 \\
$R_{9}=\langle 1^2,2,4,7,8,9     \rangle$    & 0 & 5 \\
$R_{10}=\langle 1,2^2,4^2,8,9,10,13\rangle$    & 0 & 6 \\
\hline
\hline
\end{tabular}
\end{center}
\noindent Table IX: List of type-B $D$-flat directions that are
$F$-flat to all orders for the model AF1. The dimension of the
direction, after
cancellation of the Fayet Iliopoulos term, is indicated in the second
column. The third column gives the number of non-anomalous $U(1)$'s
broken along the flat direction.

\begin{center}
\begin{tabular}{|c|rrrrrrrrrrrrrr|r|r|r|}
\hline\hline
NA         &&&&&&&&&&&&&&&&&\\
Singlet      		  &$Q_A$&$Q_1$&$Q_2$&$Q_3$&$Q_4$&$Q_5$&$Q_6$
                  &$Q_7$&$Q_8$&$Q_9$&$Q_{10}$&$Q_{11}$&$Q_{12}$&$Q_{13}$
                  &$Q_Y$&$\hat{Q}$&$Q'$\\
\hline\hline
$S_1$$\surd$&40&   0&   0&   0&   0&   0&  $-$8&  16&   0&  $-$4& 336&
     $-$40&$-$248& 136&0&0&0\\
 $S_2$$\surd$&20&   0&   0&   0&   0&   0&   2&  18& $-$10& $-$32&$-$114&
   $-$20&$-$124&  68&$-$1/2&0&0\\
 $S_3$$\surd$&16&   0&   4&   0&   0&   0&   0& $-$44&   0& $-$32& $-$64&
$-$64& 112& $-$48&$-$1/2&0&4\\
 $S_4$$\surd$&16&   0&  $-$4&   0&   0&   0&   0& $-$44&   0& $-$32&
$-$64& $-$64& 112& $-$48&$-$1/2&0&$-$4\\
 $S_5$$\surd$&12&   0&   0&   0&   0&   0&   0& $-$22&  12& $-$26&   8&
8&$-$332& $-$36&$-$1/2&0&0\\
 $S_6$$\surd$&12&   0&   0&   0&   0&   0& $-$12&   2&   0& $-$22&
$-$44&$-$232&$-$124& $-$36&$-$1/2&0&0\\
\hline
\end{tabular}

\begin{tabular}{|c|rrrrrrrrrrrrrr|r|r|r|}
\hline
NA         &&&&&&&&&&&&&&&&&\\
Singlet      		  &$Q_A$&$Q_1$&$Q_2$&$Q_3$&$Q_4$&$Q_5$&$Q_6$
                  &$Q_7$&$Q_8$&$Q_9$&$Q_{10}$&$Q_{11}$&$Q_{12}$&$Q_{13}$
                  &$Q_Y$&$\hat{Q}$&$Q'$\\
\hline\hline
 $S_{44}$&$-$20&   0&   0&   0&   0&   0&  $-$2&  70&  22&   0& 110& 204&
332& $-$68&1&0&0\\
 $S_{45}$&$-$20&   0&   0&   0&   0&  $-$4&  $-$8&  16&  12& $-$14&$-$140&
48& $-$84& $-$68&0&0&0\\
 $S_{46}$&$-$20&   0&   0&   0&  $-$4&   0&  $-$2& $-$18&  10&  32& 114&
20& 124& $-$68&1/2&0&0\\
 $S_{47}$&$-$20&   0&   0&   2&   2&   2&   6&  32&  12& $-$18&  24&  24&
276&  60&0&$-$64&0\\
 $S_{48}$&$-$20&   0&   0&   2&  $-$2&   2&   6& $-$56&   0&  14&
28&$-$160&  68&  60&$-$1/2&$-$64&0\\
 $S_{49}$&$-$20&   0&   0&  $-$2&   2&  $-$2&   6& $-$56&   0&  14&
28&$-$160&  68&  60&0&$-$64&0\\
\hline\hline
\end{tabular}
\end{center}
\noindent Table X: List of non-Abelian singlets for model CHL1, with 
$\hat{Q}=Q_A-\overline{Q}$, $Y$ and $Q'$ as defined in
eqs.~(\ref{chl1qb}), (\ref{chl1y}) and (\ref{chl1qp}) respectively. 

\begin{center}
\begin{tabular}{|l|r||l|r|}
\hline\hline
BASIS      &   $\hat{Q}$ & &$\hat{Q}$\\
\hline\hline
$M_{1}=\langle 1,\ov{1}\rangle$ &0 &
	$M_{12}=\langle 77,62,21^2,20^2\rangle$ &0 \\
$M_{2}=\langle 54,39,1\rangle$ &0 &
	$M_{13}=\langle 77,61,22,21^2,20\rangle$ &0 \\
$M_{3}=\langle 54,43,7\rangle$ &0 & 
	$M_{14}=\langle 45,42,31,29^2,28,\ov{1}\rangle$ &$-$256 \\
$M_{4}=\langle 31,21,20,7\rangle$ &0 & 
	$M_{15}=\langle 43,39,31,30,29^2,20,\ov{1}\rangle$ &$-$256 \\
$M_{5}=\langle 28,22,21,7\rangle$ &0 & 
	$M_{16}=\langle 63,45,39,38,31,30,29,28\rangle$ &$-$256 \\
$M_{6}=\langle 74,56,52,7\rangle$ &0 & 
	$M_{17}=\langle 55,47,45,39,31,30,29,28\rangle$ &$-$256 \\
$M_{7}=\langle 78,75,21,20\rangle$ &0 & 
	$M_{18}=\langle 55,45,39,38,31,29,28,\ov{1}\rangle$ &$-$256 \\
$M_{8}=\langle 77,76,22,21\rangle$ &0 & 
	$M_{19}=\langle 49,42,39,22,21^2,20,14,1\rangle$ &0 \\	
$M_{9}=\langle 77,30,21,20,1\rangle$ &0 & 
	$M_{20}=\langle 49,43,42,34,31,30,29,20,\ov{1}\rangle$ &$-$256 \\
$M_{10}=\langle 75,30,22,21,1\rangle$ &0 &
	$M_{21}=\langle 74^2,66^2,56^2,49,45,39^3,38^4,14\rangle$ &$-$256\\
$M_{11}=\langle 74,45,42,21,1\rangle$ &0 &
	$-----------------------$ &$---$ \\
\hline
\hline
\end{tabular}
\end{center}
\noindent Table XI: Basis of the moduli space of
non-anomalous $D$-flat directions of model CHL1, with $Y=0$.

\begin{center}
\begin{tabular}{|l|r||l|r|}
\hline\hline
BASIS      &   $\hat{Q}$   &   & $\hat{Q}$ \\
\hline\hline
$M_{1}=\langle 1,\ov{1}\rangle$ &0 &
	$M_{38}=\langle 22,21,5,\ov{2}\rangle$ &0 \\
$M_{2}=\langle 2,\ov{2}\rangle$ &0 &
	$M_{39}=\langle 60,10,6,\ov{5}\rangle$ &0 \\
$M_{3}=\langle 3,\ov{3}\rangle$ &0 & 
	$M_{40}=\langle 24,17,6,\ov{2}\rangle$ &0 \\
$M_{4}=\langle 4,\ov{4}\rangle$ &0 & 
	$M_{41}=\langle 23,18,6,\ov{2}\rangle$ &0 \\
$M_{5}=\langle 5,\ov{5}\rangle$ &0 & 
	$M_{42}=\langle 21,20,6,\ov{2}\rangle$ &0 \\
$M_{6}=\langle 6,\ov{6}\rangle$ &0 & 
	$M_{43}=\langle 57,33,15,7\rangle$ &0 \\
$M_{7}=\langle 68,32,1\rangle$ &0 & 
	$M_{44}=\langle 51,37,15,7\rangle$ &0 \\
\hline
\hline
\end{tabular}
\end{center}
\noindent Table XII: Basis of the moduli space of
non-anomalous $D$-flat directions of model CHL1.

\begin{center}
\begin{tabular}{|c|r rrrrrrr|r|r|}
\hline\hline
NA          & & & & & & & & & & \\
Singlet       &$Q_A$&$Q_1$&$Q_2$&$Q_3$&$Q_4$&$Q_5$&$Q_6$
                  &$Q_7$
                  &$\hat{Q}$&$Q'$\\
\hline\hline
$S_1$ $\surd$&   40&  0&  0& $-$4&  4&  8&  8& $-$8&   0&0\\
$S_2$        &   32& $-$4&  0&  4&  4&$-$16&  0& $-$8&   0&0\\
$S_3$        &    8&  6&  4&  0&  0& 24&  0&$-$12&   0&0\\
$S_4$        &    8&  6& $-$4&  0&  0& 24&  0&$-$12&   0&0\\
$S_5$        &    0&  0&  8&  0&  0&  0& 16&  0&   0&0\\
$S_6$        &    0&  0& $-$8&  0&  0&  0& 16&  0&   0&0\\ 
$S_{ 7}$     &    0& $-$4&  0&  0&  0&  0&  0& 24&   0&0\\ 
$S_{ 8}$     &  $-$40&  2&  4& $-$4& $-$4& $-$8&  0& $-$4&   0&0\\ 
$S_{ 9}$     &  $-$40&  2& $-$4& $-$4& $-$4& $-$8&  0& $-$4&   0&0\\ 
$S_{10}$     &  $-$40&  0&  0&  4& $-$4& $-$8&  8&  8&   0&0\\ 
$S_{11}$     &  $-$40&  0&  0&  0&  8& $-$8&  8&  8&$-$112&48\\ 
$S_{12}$     &  $-$16&  0&  0&  0&  0&$-$48&  0&  0&   0&48\\ 
$S_{13}$     &   $-$8&  2&  4&  0&  0&$-$24&  0& 12&   0&48\\ 
$S_{14}$     &   $-$8&  2& $-$4&  0&  0&$-$24&  0& 12&    0&48\\ 
$S_{15}$     &   $-$8& $-$2&  4&  0&  0&$-$24& 16&$-$12&   0&0\\ 
$S_{16}$     &   $-$8& $-$2&  4&  0&  0&$-$24&$-$16&$-$12&   0&0\\ 
$S_{17}$     &   $-$8& $-$2& $-$4&  0&  0&$-$24& 16&$-$12&   0&0\\ 
$S_{18}$     &   $-$8& $-$2& $-$4&  0&  0&$-$24&$-$16&$-$12&   0&0\\ 
\hline
\hline
\end{tabular}
\end{center}
\noindent Table XIII: List of non-Abelian singlet fields for model CHL2, with 
$\hat{Q}=Q_A-\overline{Q}$ as defined in eq.~(\ref{chl2qb}). Also shown is the
$Q'$ charge defined in eq.~(\ref{chl2qp}).

\vskip 1.truecm

\begin{center}
\begin{tabular}{|l|c|}
\hline\hline
BASIS      &   $\hat{Q}$  \\
\hline\hline
$M_1=\langle 1,\ov{1}\rangle$ &0 \\
$M_{2}=\langle 2,3,7,9\rangle$&0  \\
$M_{3}=\langle 2,4,7,8\rangle$ &0 \\
$M_{4}=\langle 3,6,7,16\rangle$&0 \\
$M_{5}=\langle 4,5,7,18\rangle$ &0 \\
$M_{6}=\langle 1,3,7,10,18\rangle$ &0 \\
$M_{7}=\langle 1,4,7,10,16\rangle$ &0 \\
$M_{8}=\langle 3^2,7^2,17,18\rangle$ &0 \\
$M_{9}=\langle 3,4,7^2,15,18\rangle$ &0 \\
\hline
\hline
\end{tabular}
\end{center}
\noindent Table XIV: Basis of the moduli space of
non-anomalous $D$-flat directions of model CHL2.

\begin{center}
\begin{tabular}{|c|rrrrrrrrrrrr|r|}
\hline\hline
NA        &&&&&&&&&&&&&\\ 
Singlet     &$Q_A$&$Q_1$&$Q_2$&$Q_3$&$Q_4$&$Q_5$&$Q_6$
                  &$Q_7$&$Q_8$&$Q_9$&$Q_{10}$&$Q_{11}$
                  &$\hat{Q}$\\
\hline\hline
$S_1$$\surd$&40&   0&   0&   0&   0&  $-$4&  $-$4&  $-$4&  $-$8&   4&   8&
56&0\\
$S_2$$\surd$&24&   0&   0&   0&   4&   0&   0&   8&   0&   0& $-$24& $-$24&0\\
$S_3$$\surd$&24&   4&   0&   0&   0&   0&   0&   8&   0&   0& $-$24& $-$24&0\\
$S_4$$\surd$&24&  $-$4&   0&   0&   0&   0&   0&   8&   0&   0& $-$24&
$-$24&0\\
$S_5$$\surd$&24&   0&   0&   0&  $-$4&   0&   0&   8&   0&   0& $-$24&
$-$24&0\\
$S_{6}$$\surd$&0& 4&   0&   0&   4&   0&   0&   0&   0&   0&   0&   0&0\\
\hline
\end{tabular}

\begin{tabular}{|c|rrrrrrrrrrrr|r|}
\hline
NA        &&&&&&&&&&&&&\\ 
Singlet     &$Q_A$&$Q_1$&$Q_2$&$Q_3$&$Q_4$&$Q_5$&$Q_6$
                  &$Q_7$&$Q_8$&$Q_9$&$Q_{10}$&$Q_{11}$
                  &$\hat{Q}$\\
\hline\hline
$S_{41}$&$-$48&   0&   0&   0&   0&   2&  $-$4&   0&  $-$4&   6&8&$-$48&0\\
$S_{42}$&$-$48&   0&   0&   0&   0&   0&   0&   0&   8&0&$-$16&96&$-$144\\
$S_{43}$&$-$48&   0&   0&   0&   0&   0&   4&   4&  16&   0&  16&   0&0\\
$S_{44}$&$-$48&   0&   0&   0&   0&  $-$2&  $-$4&   0&$-$4&$-$6&8& $-$48&0\\
$S_{45}$&$-$28&   2&   2&  $-$2&   2&  $-$2&   0&   8&  $-$8&2&12&$-$20&0\\
$S_{46}$&$-$28&  $-$2&  $-$2&   2&  $-$2&  $-$2&   0&   8& $-$8&2&12&$-$20&0\\
\hline
\hline
\end{tabular}
\end{center}
\noindent Table XV: List of non-Abelian singlet fields for model CHL3,
with $\hat{Q}=Q_A-\overline{Q}$ as defined in eq.~(\ref{chl3qb}).

\begin{center}
\begin{tabular}{|l|r||l|r|}
\hline\hline
BASIS      &   $\hat{Q}$   &   & $\hat{Q}$ \\
\hline\hline
$M_{1}=\langle 1,\ov{1}\rangle$ &0 &
	$M_{35}=\langle \ov{6},11,45\rangle$ &0 \\
$M_{2}=\langle 2,\ov{2}\rangle$ &0 &
	$M_{36}=\langle \ov{8},14,48\rangle$ &0 \\
$M_{3}=\langle 3,\ov{3}\rangle$ &0 & 
	$M_{37}=\langle \ov{9},18,51\rangle$ &0 \\
$M_{4}=\langle 4,\ov{4}\rangle$ &0 & 
	$M_{38}=\langle \ov{9},19,50\rangle$ &0 \\
$M_{5}=\langle 5,\ov{5}\rangle$ &0 & 
	$M_{39}=\langle \ov{6},20,49\rangle$ &0 \\
$M_{6}=\langle 6,\ov{6}\rangle$ &0 & 
	$M_{40}=\langle \ov{6},25,64\rangle$ &0 \\
$M_{7}=\langle 7,\ov{7}\rangle$ &0 & 
	$M_{41}=\langle \ov{6},26,63\rangle$ &0 \\
\hline
\hline
\end{tabular}
\end{center}
\noindent Table XVI: Basis of the moduli space of
non-anomalous $D$-flat directions of model CHL3.

\begin{center}
\begin{tabular}{|c|rrrrrrr|r|r|}
\hline\hline
NA         &&&&&&&&&\\
Singlet &$Q_A$&$Q_1$&$Q_2$&$Q_3$&$Q_4$&$Q_5$&$Q_6$
                  &$Q_Y$&$\hat{Q}$\\
\hline\hline
$S_1$ $\surd$    &16&  0&  0&$-$20&  4& 20&$-$32&$-$1/2&0\\   
$S_2$ $\surd$    &16&  4&  0&$-$12&$-$20& $-$4&$-$32&$-$1/2&0\\   
$S_3$ $\surd$    &16&  2&$-$32&  0& $-$8&$-$28&$-$32&$-$1/2&0\\   
$S_4$ $\surd$    &16& $-$2&  0& $-$4&$-$20& 32&$-$32&$-$1/2&0\\   
$S_5$ $\surd$    &16& $-$2&$-$32& $-$8& 16& $-$4&$-$32&$-$1/2&0\\   
$S_6$ $\surd$    &16& $-$4&$-$32&  8& $-$8&  8&$-$32&$-$1/2&0\\
\hline
\hline
\end{tabular}
\end{center}
\noindent Table XVII: Same as Table I but for model CHL4, with hypercharge
and $\hat{Q}=Q_A-\overline{Q}$ as defined in eqs.~(\ref{chl4y}) and
(\ref{chl4qb}) respectively.

\begin{center}
\begin{tabular}{|l|c|}
\hline\hline
BASIS      &   $\hat{Q}$  \\
\hline\hline
$M_1=\langle 10,\ov{10}\rangle$ &0 \\
$M_{2}=\langle 11,\ov{11}\rangle$&0  \\
$M_{3}=\langle 12,\ov{12}\rangle$ &0 \\
$M_{4}=\langle \ov{10},12,16\rangle$&0 \\
$M_{5}=\langle \ov{10},15,20\rangle$ &0 \\
$M_{6}=\langle 10,\ov{11},28\rangle$ &0 \\
$M_{7}=\langle 11,\ov{12},19\rangle$ &0 \\
$M_{8}=\langle \ov{12},18,20\rangle$ &0 \\
$M_{9}=\langle 10,21,29\rangle$ &0 \\
$M_{10}=\langle 12,17,29\rangle$ &0 \\
$M_{11}=\langle 15,19,29\rangle$ &0 \\
$M_{12}=\langle 16,20,29,31^2\rangle$ &0\\
\hline
\hline
\end{tabular}
\end{center}
\noindent Table XVIII: Basis of the moduli space of
non-anomalous $D$-flat directions of model CHL4, involving
$Y=0$ fields only.

\begin{center}
\begin{tabular}{|c|rrrrrrrrrrr|r|}
\hline\hline
NA       &&&&&&&&&&&&\\
Singlet    &$Q_A$&$Q_1$&$Q_2$&$Q_3$&$Q_4$&$Q_5$&$Q_6$  
                 &$Q_7$&$Q_8$&$Q_9$&$Q_{10}$
                 &$\hat{Q}_{A}$\\ 
\hline\hline
$S_1$ $\surd$   &   40&   0&   0&   0&  $-$4&  $-$4&  $-$4&  $-$8&   4&   8&  56&0\\  
$S_2$ $\surd$   &   28&   0&   0&   0&   2&   0&  $-$8&   8&  $-$2& $-$12&  20&0\\ 
$S_3$ $\surd$   &   24&   0&   0&   0&   0&   0&   8& $-$16&   0&   8&  24&0\\  
$S_4{}^({}'{}^)$ $\surd{}^({}'{}^)$   
                 &   24&   0&   0&   0&   0&   0&   8&     0& 0& $-$24& $-$24&0\\  
$S_5$ $\surd$   &   16&   0&   0&   0&  $-$2&   0&   4& $-$12&  10&  $-$8& $-$16&0\\  
$S_6$ $\surd$   &   16&   0&   0&   0&  $-$6&   0&   4& $-$12&  $-$2&  $-$8& $-$16&0\\
\hline
\hline
\end{tabular}
\end{center}
\noindent Table XIX: Same as Table I but for model CHL6, 
with $\hat{Q}=Q_A-\overline{Q}$ as defined in eq. (\ref{chl6qb}).

\begin{center}
\begin{tabular}{|l|r||l|r|}
\hline\hline
BASIS      &   $\hat{Q}$   &   & $\hat{Q}$ \\
\hline\hline
$M_{1}=\langle 1,\ov{1}\rangle$ &0 &
	$M_{22}=\langle 2,7,8,25\rangle$ &0 \\
$M_{2}=\langle 2,\ov{2}\rangle$ &0 &
	$M_{23}=\langle 2,7,12,22\rangle$ &0 \\
$M_{3}=\langle 3,\ov{3}\rangle$ &0 & 
	$M_{24}=\langle 3^2,26, 27\rangle$ &0 \\
$M_{4}=\langle 4,\ov{4}\rangle$ &0 & 
	$M_{25}=\langle 3,5,26,33\rangle$ &0 \\
$M_{5}=\langle 5,\ov{5}\rangle$ &0 & 
	$M_{26}=\langle 3,5,27,29\rangle$ &0 \\
$M_{6}=\langle 6, \ov{6}\rangle$ &0 & 
	$M_{27}=\langle 3,6,26,32\rangle$ &0 \\
$M_{7}=\langle 7, \ov{7}\rangle$ &0 & 
	$M_{28}=\langle 3,6,27,28\rangle$ &0 \\
$M_{8}=\langle 8,  \ov{8}\rangle$ &0 & 
	$M_{29}=\langle 3,\ov{5},16,22\rangle$ &0 \\	
$M_{9}=\langle 9, \ov{9}\rangle$ &0 & 
	$M_{30}=\langle 3,\ov{6},20,35\rangle$ &0 \\
$M_{10}=\langle 10, \ov{10}\rangle$ &0 &
	$M_{31}=\langle  3,\ov{5},21,34\rangle$ &0 \\
$M_{11}=\langle 11, \ov{11}\rangle$ &0 &
	$M_{32}=\langle 3,\ov{5},\ov{11},\ov{12}\rangle$&0 \\
$M_{12}=\langle 12, \ov{12}\rangle$ &0 &
	$M_{33}=\langle \ov{3},5,17,19\rangle$ &0 \\
$M_{13}=\langle 13, \ov{13}\rangle$ &0 &
	$M_{34}=\langle \ov{3},6,17,18 \rangle$ &0 \\
$M_{14}=\langle 14, \ov{14}\rangle$ &0 &
	$M_{35}=\langle  1,3,\ov{5},20,23\rangle$ &0 \\
$M_{15}=\langle 15, \ov{15}\rangle$ &0 &
	$M_{36}=\langle  1,3,\ov{6},21,23\rangle$ &0 \\
$M_{16}=\langle 1,4, \ov{7},23\rangle$ &0 &
	$M_{37}=\langle  1,8,\ov{11},17,25\rangle$ &0 \\
$M_{17}=\langle 2,3,\ov{9},22\rangle$ &0 &
	$M_{38}=\langle \ov{1},2,3,9,31\rangle$ &0 \\
$M_{18}=\langle 2,3, \ov{10},23\rangle$ &0 &
	$M_{39}=\langle  \ov{1},2,3,13,30\rangle$ &0 \\
$M_{19}=\langle 2,3, \ov{13},25\rangle$ &0 &
	$M_{40}=\langle 5,\ov{7},23,24,25\rangle$ &$-$144 \\
$M_{20}=\langle 2,5,10,25\rangle$ &0 &
	$M_{41}=\langle 4',\ov{4}\rangle$ &0 \\
$M_{21}=\langle 2,6,9,23 \rangle$ &0 &
	$M_{42}=\langle 4, \ov{4}'\rangle$ &0 \\
\hline
\hline
\end{tabular}
\end{center}
\noindent Table XX: Basis of the moduli space of
non-anomalous $D$-flat directions of model CHL6.

\begin{center}
\begin{tabular}{|c|rrrrrrrrr|r|}
\hline\hline
NA         &&&&&&&&&&\\
Singlet  &$Q_A$&$Q_1$&$Q_2$&$Q_3$&$Q_4$&$Q_5$&$Q_6$&$Q_7$&$Q_8$
                  &$\hat{Q}$\\
\hline\hline
$S_1$ $\surd$&  16& 0& 0&$-$2& 0& 0&  8& $-$4&  6& 0\\ 
$S_2{}^({}'{}^)$ $\surd$&   12& 2& 0&$-$2& 0& 4& $-$4&  0& $-$6& 0\\  
$S_3{}^({}'{}^)$ $\surd$&    4& 2& 2&$-$1& 0& 4& $-$8& $-$6&  3&  0\\ 
$S_4{}^({}'{}^)$ $\surd$&    4& 2&$-$2&$-$1& 0& 4& $-$8& $-$6&  3 &0\\  
$S_5$ $\surd$&    4&$-$2& 0& 4& 0&$-$4& $-$4&  4&  0& 32\\   
$S_6{}^({}'{}^)$ $\surd{}^({}'{}^)$&    0& 4& 0& 0& 0& 0&  0&  0&  0& 0\\
\hline
\end{tabular}
\end{center}

\begin{center}
\begin{tabular}{|c|rrrrrrrrr|r|}
\hline
NA        &&&&&&&&&&\\
Singlet  &$Q_A$&$Q_1$&$Q_2$&$Q_3$&$Q_4$&$Q_5$&$Q_6$&$Q_7$&$Q_8$
                  &$\hat{Q}$\\
\hline\hline
$S_{41}$     &  $-$12& 0& 0&$-$4& 2&$-$4&  4&  0& $-$6& $-$32\\
$S_{42}$     &  $-$12& 0&$-$2& 3&$-$2& 4&  0&  6& $-$3& 0\\ 
$S_{43}$     &  $-$12& 0&$-$2& 1& 0& 4& 16& $-$2& $-$3& $-$32\\ 
$S_{44}$     &  $-$12& 0&$-$2&$-$1& 0& 4&  8& 10&  3& 0\\   
$S_{45}$     &  $-$8& 2& 0& 2&$-$2&$-$8& $-$8&  0&  0& 0\\   
$S_{46}$     &  $-$8& 2& 0& 0& 2&$-$8&  0&  4& $-$6& 0\\   
$S_{47}$     &  $-$8& 2& 0& 0& 0&$-$8&  8& $-$8&  0& $-$32\\   
$S_{48}$     &  $-$8& 2& 0&$-$2& 0&$-$8&  0&  4&  6& 0\\   
%$S_{49}$    &  $-$8& 0& 2&$-$1& 4& 0&  4& $-$2&  3& 0\\   
%$S_{50}$    &  $-$8& 0& 2&$-$1&$-$4& 0&  4& $-$2&  3 &$-$32\\   
%$S_{51}$    &  $-$8& 0&$-$2&$-$1& 4& 0&  4& $-$2&  3 &0\\   
%$S_{52}$    &  $-$8& 0&$-$2&$-$1&$-$4& 0&  4& $-$2&  3 &$-$32\\   
$S_{49}$     &  $-$8&$-$2& 0& 2&$-$2& 8& $-$8&  0&  0 &0\\   
$S_{50}$     &  $-$8&$-$2& 0& 0& 2& 8&  0&  4& $-$6& 0\\   
$S_{51}$     &  $-$8&$-$2& 0& 0& 0& 8&  8& $-$8&  0& $-$32\\   
$S_{52}$     &  $-$8&$-$2& 0&$-$2& 0& 8&  0&  4&  6& $-$32\\   
$S_{53}$     &  $-$4& 0& 2& 1&$-$2&$-$4&$-$16& $-$6&  3 &0\\   
$S_{54}$     &  $-$4& 0& 2&$-$1& 0&$-$4&  0&$-$14&  3& $-$32\\   
$S_{55}$     &  $-$4& 0& 2&$-$3& 0&$-$4& $-$8& $-$2&  9 &0\\   
$S_{56}$     &  $-$4& 0& 0& 4& 2& 4& $-$4&  8& $-$6&  32\\  
$S_{57}$     &  $-$4& 0& 0& 2& 4& 4& 12&  0& $-$6& 0\\   
$S_{58}$     &  $-$4& 0& 0& 2&$-$4& 4& 12&  0& $-$6& $-$32\\   
$S_{59}$     &  $-$4& 0& 0& 0& 4& 4&  4& 12&  0& 32\\    
$S_{60}$     &  $-$4& 0& 0& 0&$-$4& 4&  4& 12&  0& 0\\   
$S_{61}$     &  $-$4& 0& 0&$-$2&$-$2& 4& 20&  4&  0&  $-$32\\ 
$S_{62}$     &  $-$4& 0&$-$2& 1&$-$2&$-$4&$-$16& $-$6&  3 &0\\   
$S_{63}$     &  $-$4& 0&$-$2&$-$1& 0&$-$4&  0&$-$14&  3& $-$32\\   
$S_{64}$     &  $-$4& 0&$-$2&$-$3& 0&$-$4& $-$8& $-$2&  9 &0\\
\hline
\hline
\end{tabular}
\end{center}
\noindent Table XXI: List of non-Abelian singlets for model CHL7, 
with $\hat{Q}=Q_A-\overline{Q}$ as defined in eq. (\ref{chl7qb}).

\begin{center}
\begin{tabular}{|l|r||l|r|}
\hline\hline
BASIS      &   $\hat{Q}$   &   & $\hat{Q}$ \\
\hline\hline
$M_{1}=\langle 1,\ov{1}\rangle$ &0 &
	$M_{35}=\langle 8,35,41\rangle$ &$-$32 \\
$M_{2}=\langle 2,\ov{2}\rangle$ &0 &
	$M_{36}=\langle 9,58,64\rangle$ &0 \\
$M_{3}=\langle 3,\ov{3}\rangle$ &0 & 
	$M_{37}=\langle 9,60,63\rangle$ &0 \\
$M_{4}=\langle 4,\ov{4}\rangle$ &0 & 
	$M_{38}=\langle 9,61,62\rangle$ &0 \\
$M_{5}=\langle 5,\ov{5}\rangle$ &0 & 
	$M_{39}=\langle 10,26,41\rangle$ &0 \\
$M_{6}=\langle 6,\ov{6}\rangle$ &0 & 
	$M_{40}=\langle 10,57,64\rangle$ &0 \\
$M_{7}=\langle 7,\ov{7}\rangle$ &0 & 
	$M_{41}=\langle 10,59,63\rangle$ &0 \\
\hline
\hline
\end{tabular}
\end{center}
\noindent Table XXII: Basis of the moduli space of
non-anomalous $D$-flat directions of model CHL7.

\end{document}